# Establishing baselines for generative discovery of inorganic crystals

Nathan J. Szymanski[1] and Christopher J. Bartel[1,*]

## Abstract

Generative artificial intelligence offers a promising avenue for materials discovery, yet its advantages over traditional methods remain unclear. In this work, we introduce and benchmark two baseline approaches – random enumeration of charge-balanced prototypes and data-driven ion exchange of known compounds – against four generative techniques based on diffusion models, variational autoencoders, and large language models. Our results show that established methods such as ion exchange are better at generating novel materials that are stable, although many of these closely resemble known compounds. In contrast, generative models excel at proposing novel structural frameworks and, when sufficient training data is available, can more effectively target properties such as electronic band gap and bulk modulus. To enhance the performance of both the baseline and generative approaches, we implement a post-generation screening step in which all proposed structures are passed through stability and property filters from pre-trained machine learning models including universal interatomic potentials. This low-cost filtering step leads to substantial improvement in the success rates of all methods, remains computationally efficient, and ultimately provides a practical pathway toward more effective generative strategies for materials discovery. By establishing baselines for comparison, this work highlights opportunities for continued advancement of generative models, especially for the targeted generation of novel materials that are thermodynamically stable.

[1] University of Minnesota, Department of Chemical Engineering and Materials Science, Minneapolis, MN, USA 55455

[*] Correspondence to cbartel@umn.edu

## Introduction

The discovery of new materials has long been a cornerstone of technological progress, driving many of the innovations that shape modern society.[1] Breakthroughs in layered and Li-rich battery cathodes, for example, have enabled the widespread adoption of portable electronics and electric vehicles.[2] Transparent conducting oxides such as indium tin oxide (ITO) and indium gallium zinc oxide (IGZO) have been critical for the development of touch screens, solar cells, and flat-panel displays.[3] Similarly, the discovery of cuprate superconductors in the 1980s reignited interest in high-temperature superconductivity, which remains the subject of extensive research.[4] These examples highlight the role of materials discovery in advancing transformative technologies and underscore the need for innovative approaches to accelerate future breakthroughs.

The recent emergence of generative artificial intelligence (AI) offers a promising route for designing new materials, particularly inorganic crystals.[5,6] Early efforts focused on generative adversarial networks (GANs)[7–9] and variational autoencoders (VAEs),[10–14] while more recent developments include large language models (LLMs)[15–19], diffusion-based techniques,[20–24] normalizing flows,[25–27] and geodesic random walks.[28] These models are often trained on computed materials from open databases such as the Materials Project[29] to generate thermodynamically stable structures, with some also conditioned on specific properties for application-driven campaigns.

There exists a growing number of successes in generative AI for materials design, with validation provided by *ab-initio* calculations and experimental synthesis. For example, text-based models such as Chemeleon[30] have leveraged contrastive learning to align crystal structures with natural language descriptions, enabling composition- and structure-conditioned generation in a variety of chemical spaces – most notably achieving successful phase prediction in the Li-P-S-Cl system relevant to solid-state batteries. Other models such as FlowLLM[26] combine language representations with Riemannian flow matching to refine generated structures, increasing their stability rate threefold. Among diffusion models, MatterGen[22] has emerged as a particularly effective method capable of generating materials with targeted chemistry, symmetry, and functional properties – even leading to the synthesis of an AI-generated compound, $TaCr_2O_6$, whose experimentally measured bulk modulus was within 20% of the predicted value.

Despite these recent successes, it remains difficult to systematically assess the performance of different generative models in a consistent fashion. Tools like *matbench-genmetrics* provide important frameworks and metrics for evaluating the validity of structures proposed by generative models,[31] while *matbench-discovery* addresses the challenge of benchmarking stability predictions made by machine learning (ML) models and interatomic potentials.[32] Yet, the extent to which generative models outperform established methods, such as ion exchange or high-throughput screening, is not yet fully understood. Baselines are therefore essential to clarify where these models offer the greatest advantages – whether in producing stable materials, generating novel structures, or achieving targeted properties – and to identify their limitations. Such benchmarks are key to integrating generative models into existing workflows and driving tangible progress in materials discovery.

In this work, we establish two baseline methods for the generation of inorganic crystals: random enumeration of charge-balanced chemical formulae in structure prototypes sourced from the AFLOW database,[33,34] and ion exchange performed on stable compounds with desired properties from the Materials Project.[29] These methods are benchmarked against four generative models – CrystaLLM,[15] FTCP,[12] CDVAE,[13] and MatterGen[22] – for the generation of (1) materials that are stable and novel, (2) materials with a band gap near 3 eV, and (3) materials with high bulk modulus. We also integrate two graph neural networks, CHGNet[35] and CGCNN,[36] to filter and retain generated materials predicted to be stable or exhibit desired properties. This evaluation sheds light on the comparative strengths and weaknesses of traditional and generative approaches to materials discovery, while also providing a set of baselines against which future generative models can be benchmarked.

## Methods

### Random enumeration

In this method, we randomly paired a known structure prototype with a set of elements also chosen at random. The prototypes were drawn from 1,783 structures listed in the Encyclopedia of Crystallographic Prototypes (AFLOW).[33,34] Compositional sets were created using three to five elements, forming ternary to quinary phases. Binary phases were excluded because they have been extensively studied already, leaving little opportunity for novel materials discovery. For each structure-composition pair, we assigned the elements to specific prototype sites based on the given chemical formula. For example, randomly selecting the prototype "A2BC4_cF56_227_c_b_e-001" (a normal spinel in the AFLOW prototype library) and a composition of Mn-Fe-S would yield six spinel structures by exploring all possible arrangements of Mn, Fe, and S on the *A*/*B*/*C* sites. Charge balance is then assessed using common oxidation states provided by *pymatgen*.[37] If charge balance is plausible, the structure of the prototype is decorated accordingly, and the resulting materials undergo further evaluation using density functional theory (DFT). In the previous example, only $Mn_2FeS_4$ and $Fe_2MnS_4$ would be retained of the six enumerated structures. We note that while this method can produce new compositions, it will never produce an entirely new structure given its reliance on established prototypes. Our use of common oxidation states further limits the breadth of compositions that may be generated, though this requirement may be relaxed by the user if more exotic materials are desired.

**Data-driven ion exchange**

The Materials Project (MP) contains DFT-calculated properties for ~153,000 compounds, providing a solid foundation for materials discovery campaigns.[29] Starting from these materials, we leveraged the data-mined substitution prediction (DMSP) algorithm[38] implemented in *pymatgen*[37] to replace one or more ions of a given compound to yield new hypothetical materials. The substitutions are guided by conditional probabilities, $p_{\mathrm{DMSP}}$, which quantify the likelihood of substituting one ion for another while retaining the original crystal structure. These probabilities are derived from a probabilistic model trained on the Inorganic Crystal Structure Database (ICSD), an experimental database of known crystal structures.[39] Substitutions were performed for pairs of species with $p_{\mathrm{DMSP}} > 0.001$ (the default value in *pymatgen*) to balance the tradeoff between generating novel substitutions and maintaining structural plausibility. Multiple substitutions were allowed per material, enabling both single and multi-site exchanges. For example, starting from $CaTiO_3$ yields substitutions like $SrTiO_3$ and $SrZrO_3$ (single-site) in addition to $SrTiS_3$ and $SrZrS_3$ (multi-site). DMSP was applied in two different modes: 1) to generate stable materials and 2) to generate materials with desired properties. For the first task, we randomly extracted stable parent structures from MP and substituted at least one ion with a species not already present in the original composition. To generate materials with desired properties, we selected materials having specified target values of that property (*e.g.*, band gaps near 3 eV) and performed ion substitution in a similar fashion to generate new materials. As with random enumeration, data-driven ion exchange can only produce materials with novel compositions, whereas the structures will always be based on known materials.

## Generative modeling

Four models were assessed in this work: CrystaLLM, FTCP, CDVAE, and MatterGen. CrystaLLM[15] is a transformer-based large-language model designed to learn from tokenized representations of crystallographic information files (CIFs). We evaluated three pre-trained versions of this model (available at github.com/lantunes/CrystaLLM): two were trained on the MP-20 dataset,[13] which contains 45,231 stable structures from the Materials Project,[29] using different model sizes (denoted *small* and *large*), and one was trained on a larger dataset of 2.3 million structures. These models require only a chemical formula as input, from which a CIF file is generated. All compositions fed to each instance of CrystaLLM were obtained through random enumeration of charge-balanced formulae. For consistent comparison with other models evaluated in this work, results shown in the main text (**Figures 1**, **2**) correspond to the *large* model trained on MP-20; results from the other two versions are provided in the Supplementary Information.

FTCP[12] encodes materials using real-space features (lattice vectors, one-hot encoded element vectors, site coordinates, and occupancies) and reciprocal-space features derived from a Fourier transform of elemental property vectors. Two FTCP-based autoencoders were trained on the MP-20 dataset: one conditioned on formation energy and electronic band gap and another conditioned on formation energy and bulk modulus. Due to the limited availability of elastic property data, the latter autoencoder was trained and validated on a subset of MP-20 containing 9,361 materials. Materials were then generated by randomly sampling points in the latent space nearby known materials from the training set. Default hyperparameters supplied at https://github.com/PV-Lab/FTCP were used for the training and generation steps.

CDVAE[13] combines a variational autoencoder (VAE) with a diffusion model to generate new materials. Sampling from the latent space predicts composition, lattice vectors, and the number of atoms in the unit cell, which are used to randomly initialize structures. The diffusion component of the model then "de-noises" these random structures by iteratively perturbing atoms toward equilibrium positions. We trained CDVAE on MP-20 without any conditioning of its latent space, allowing it to be used only for the generation of stable materials. During the generation step of each model, we did not place any constraints on the elements or symmetries that may be created. Default hyperparameters supplied at https://github.com/txie-93/cdvae were used for the training and generation steps.

MatterGen[22] is a more recent diffusion model designed to create stable materials by jointly denoising atom types, coordinates, and lattice parameters through an equivariant score network. MatterGen can be fine-tuned for targeted properties but was used here in its base configuration (available at https://github.com/microsoft/mattergen) for unconstrained generation, re-trained on the MP-20 dataset for consistent comparison with the other models evaluated in this work. This model was used to generated structures in batches of 128 using the provided command-line tool. A similar procedure was used to generate materials from a MatterGen model that was pre-trained on a much larger dataset (Alex-MP-20), with results provided in the Supplementary Information.

**Novelty assessment**

A material was considered novel if it was not already present in MP (as of June 2025), thus ensuring it was excluded from the training data used for the generative models evaluated here. To determine this, we queried all entries in MP with the same composition as a proposed structure. For each resulting entry, structural similarity was assessed using the *StructureMatcher* tool in *pymatgen*,[37] which compares structures based on their lattice parameters, atomic positions, and symmetry. We used loose tolerance parameters for this comparison to account for slight variations in computed structures. This included a lattice parameter tolerance of 0.25 Å, a site tolerance defined as 40% of the average free length per atom, and an angular tolerance of 10°. If no matching composition or structure was identified in the MP database, the material was classified as novel.

**Density functional theory calculations**

Novel materials obtained from each generation approach described above were relaxed using the PBE exchange-correlation functional[40] within the projector augmented wave (PAW) method as implemented in the Vienna *Ab Initio* Simulation Package (VASP).[41,42] These calculations used a plane-wave cutoff energy of 520 eV with a Γ-centered k-point grid spacing of 0.22 $Å^{-1}$. Convergence criteria were set to $10^{-6}$ eV for the electronic optimizations and 0.03 eV/Å for the ionic relaxations. This relatively loose convergence criterion was determined to be sufficient for evaluating stability in high throughput, though a tighter criterion should be used if the objective is to precisely determine equilibrium atomic positions (Supplementary Figure 1). Symmetry was turned off to ensure accurate treatment of distortions, and spin polarization was included for materials containing potentially magnetic elements (Ti, V, Cr, Mn, Fe, Co, Ni, Cu,

Nb, Mo, Tc, Ru, Rh, Pd, Ag, Ta, W, Re, Os, Ir, Pt, Ce, Eu, Pr, Nd, Pm, Sm, Gd, Tb, Dy, Ho, Er, Tm, Yb, Np, and U). These elements were chosen to be consistent with the Materials Project. All magnetic moments were initialized in a ferromagnetic configuration for all such compounds. For materials containing $3d$ transition metals, Hubbard U corrections (+U) were applied to account for on-site Coulomb interactions following the conventions used in the Materials Project: 4.0 eV for Mn, 3.9 eV for Fe, 3.7 eV for Co, 6.2 eV for Ni, and 5.3 eV for Cu.[29]

Thermodynamic stability with respect to all known competing phases in MP was evaluated using the decomposition energy ($\Delta E_{\mathrm{d}}$).[43] For unstable materials with $\Delta E_{\mathrm{d}} > 0$, this measure is equivalent to the energy above the convex hull ($E_{\mathrm{hull}}$). It quantifies the energy difference between the proposed material and the lowest energy combination of competing phases. For stable materials with $\Delta E_{\mathrm{d}} \leq 0$, the decomposition energy is the energy by which the proposed material lies below the existing convex hull (if the proposed material were not included in its construction). Total energies acquired from DFT calculations were transformed into formation energies ($\Delta E_{\mathrm{f}}$) using the *MaterialsProject2020* compatibility scheme, which accounts for GGA/GGA+U mixing and implements elemental reference energy corrections as described in previous work.[44] Competing phases were identified by constructing a phase diagram for each chemical system using the *PhaseDiagram* module from *pymatgen*.[37] These phase diagrams included all entries from MP (as of June 2025) as well as the entries generated from each AI model or baseline approach, allowing for an evaluation of stability against both known and hypothetical phases.

Electronic band gaps were computed by analyzing the eigenvalue band properties obtained from VASP calculations using *pymatgen*,[37] with the band gap defined as the energy difference between the valence band maximum and conduction band minimum. The bulk modulus of each structure was computed by fitting a Birch-Murnaghan equation of state[45] to relaxed (but fixed volume) total energy calculations performed at seven volumes ranging from 97% to 103% of the equilibrium volume. These volumes were generated by isotropically scaling the lattice vectors of the relaxed equilibrium structures. The equilibrium bulk modulus and its pressure derivative were extracted from the fit, providing a measure of each material's resistance to volumetric deformation.

**Machine learning filtering**

Universal machine learning interatomic potentials (uMLIPs) offer an efficient way to screen large numbers of candidate materials obtained from any of the methods described above, enabling the rapid identification of promising structures prior to more computationally expensive DFT calculations.[46] Here, we used CHGNet[35] to compute the internal (0 K) energies of all candidate materials, which were then compared against the convex hulls in MP to assess thermodynamic stability. Each structure was relaxed using the Atomic Simulation Environment (ASE)[47] with CHGNet-based force fields, ensuring the forces acting on its atoms converged below 0.1 eV/Å. This relatively loose convergence criterion was chosen to improve computational efficiency for the task of high-throughput screening. CHGNet was trained on GGA and GGA+U calculations with *MaterialsProject2020Compatibility* corrections applied,[44] enabling direct comparison with MP energies. Thermodynamic stability was evaluated by constructing phase diagrams for the relevant chemical systems, incorporating all MP entries (as of June 2025) alongside generated candidate structures. Materials predicted to be stable within these phase diagrams were passed to DFT calculations for further validation.

For property-specific screening, we leveraged two CGCNN[36] models: one trained on 16,458 DFT-calculated band gaps and the other on 2,041 bulk moduli from MP. When generating materials with a high bulk modulus, we applied an acceptance criterion of CGCNN-predicted bulk moduli exceeding 200 GPa. Analogously, when generating materials with a band gap near 3 eV, we selected candidates with CGCNN-predicted band gaps in the range of 2.8 to 3.2 eV.

## Results

### Stability and novelty

Each method described above was used to generate a sufficient number of candidates such that 500 novel materials – *i.e.*, ones that are not already present in the Materials Project (MP) – could be selected and passed to DFT calculations. The average time required by each method to generate these materials is listed in Supplementary Table 1. Their decomposition energies ($\Delta E_{\mathrm{d}}$), computed relative to the convex hull defined by MP entries, are shown in **Figure 1**. Random enumeration of known structure prototypes with charge-balanced chemical formulae results in a wide range of decomposition energies with a median value ($\Delta E_{\mathrm{d}}^{\mathrm{med}}$) of 409 meV/atom. Only 1.4% of the novel materials generated with this method are thermodynamically stable ($\Delta E_{\mathrm{d}} \leq 0$). In contrast, a much higher stability rate can be achieved by leveraging analogies to known materials. Ion exchange results in a much tighter distribution of novel materials ($\Delta E_{\mathrm{d}}^{\mathrm{med}} = 85$ meV/atom) close to the hull, with 9.2% of them being stable.

The high stability rate of ion exchange is impressive but perhaps unsurprising, given its proven efficacy in discovering new materials using high-throughput calculations performed over the past decade.[48–53] This is especially true when leveraging known structure prototypes that can host a wide variety of compositions. For example, prior work has identified several hundred stable compositions in the perovskite structure through ion exchange of known materials.[54–56] Similar approaches have also been applied to successfully uncover stable compositions in the spinel and delafossite structures,[57,58] reinforcing the effectiveness of ion exchange for materials discovery.

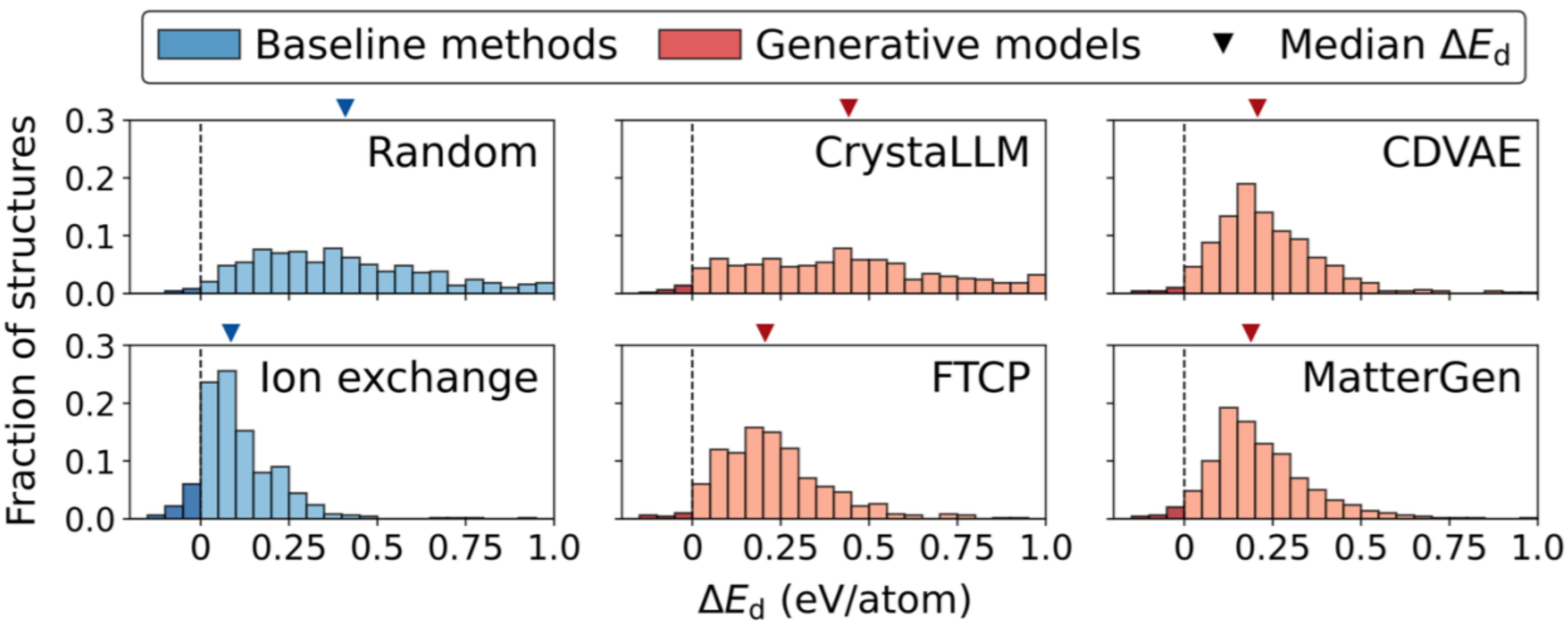

**Figure 1:** Histograms showing DFT-computed decomposition energies ($\Delta E_d$) of novel materials (not already present in the Materials Project) generated by two baseline methods and four generative models. For each of these six approaches, 500 materials were considered. The left column (blue) contains results from the baseline methods: random enumeration and ion exchange. The right columns (red) contain results from the generative models: CrystaLLM,[15] FTCP,[12] CDVAE,[13] and MatterGen.[22] Triangular markers indicate median decomposition energies.

Among the generative models, CrystaLLM produced novel materials with the widest range of energies ($\Delta E_d^{med}$ = 442 meV/atom) but a stability rate (2.4%) that is second only to MatterGen (3.0%). While both remain well below the stability rate of ion exchange (9.2%), the generative models provide unique flexibility in terms of capacity and training data. For example, the CrystaLLM model used in **Figure 1** had 200 million parameters and outperformed a smaller model variant with only 25 million parameters and a lower stability rate of 1.6% (Supplementary Figure 2). Expanding the training set also improves performance, with CrystaLLM achieving a stability rate of 2.8% when trained on ~2.3 million structures (Supplementary Figure 3). Similar improvements are possible with MatterGen, whose stability rate increases from 3.0% to 5.4% when trained on structures from the Alexandria database[59] in addition to those from MP-20 (Supplementary Figure 4).

As both CDVAE and MatterGen are diffusion models, they result in a similar distribution of energies with $\Delta E_{\mathrm{d}}^{\mathrm{med}}$ of 207 and 188 meV/atom, respectively. These two models also produce comparable energy distributions to FTCP, a variational autoencoder, which yields $\Delta E_{\mathrm{d}}^{\mathrm{med}} = 205$ meV/atom and a moderate stability rate of 2.0%. Given the recent success of diffusion models, the competitiveness of FTCP (which is a VAE) is somewhat unexpected. We attribute this to its latent space sampling strategy, which biases generation toward materials that are structurally similar to known stable compounds. This approach enhances the model's stability rate, similar to the observed benefits of template-based strategies (such as ion exchange) in the baseline methods. However, it also comes at the cost of reduced novelty – 1,309 materials were generated from FTCP before obtaining 500 materials not already present in MP (a novelty rate of ~38.2%).

In this work, novelty is defined as a material being absent from MP. While this does not necessarily indicate the material has never been synthesized or is absent from all computational databases, it signifies that the material was not used for training of the generative models or as a template for ion exchange. All materials shown in **Figure 1** meet this definition of novelty. However, the total number of generated materials required to obtain 500 novel ones varied across methods. The rate at which novel materials were produced by each method is listed in **Table 1**.

Between the two baseline methods, random enumeration yields a much higher novelty rate (98.6%) than ion exchange (72.4%). This reflects the unconstrained nature of random enumeration, which leads to the sampling of many previously unexplored chemical compositions. In contrast, our approach to ion exchange closely reflects traditional screening efforts,[60] and is therefore more likely to reproduce materials already present in computational databases such as MP. However, the use of ion exchange also comes with the benefit of generating more stable materials, resulting in a higher stability rate (9.2%) than random enumeration (1.4%).

**Table 1:** Stability and novelty rates of materials generated from each method. $\Delta E_{\mathrm{d}}^{\mathrm{med}}$ is the median decomposition energy of generated materials not already present in the Materials Project. Also listed is the prototype novelty rate, defined as the percentage of proposed materials whose structures cannot be indexed to a known prototype in the AFLOW database, and the stability rate of materials in these novel prototypes. The bold value in each column denotes the highest rate achieved among all methods. Statistics are based on 500 novel materials generated by each method.

| Method | $\Delta E_{\mathrm{d}}^{\mathrm{med}}$ (meV/atom) | Stability rate | Novelty rate | Novel prototype rate | Novel prototype stability rate |
|---|---|---|---|---|---|
| Random | 409 | 1.4% | **98.6%** | 0% | 0% |
| Ion exchange | **85** | **9.2%** | 72.4% | 0% | 0% |
| CrystaLLM | 442 | 2.4% | 98.2% | 1.0% | 0% |
| CDVAE | 207 | 1.8% | 96.0% | **8.2%** | 0% |
| FTCP | 205 | 2.0% | 38.2% | 1.8% | 0% |
| MatterGen | 188 | 3.0% | 91.8% | 7.2% | 0% |

Three of the generative models – CrystaLLM, CDVAE, and MatterGen – exhibit high novelty rates > 90%. In contrast, only 38.2% of the materials generated by FTCP are novel. This result is consistent with FTCP's strategy of sampling around known materials in its latent space, a factor that likely also contributes to its higher stability rate. To assess the impact of the sampling strategy used by FTCP, we generated several new sets of materials at iteratively greater distances from known materials in its latent space. The results, shown in Supplementary Table 2 and Supplementary Figure 5, demonstrate that sampling further away from known materials leads to higher novelty rates (reaching 95%) but also lower stability rates ($\leq$ 1%). The clear inverse correlation between these two metrics underscores the tradeoff that exists between stability and novelty during materials discovery campaigns.

To more broadly assess novelty, we examined the fraction of generated materials absent from two additional sources: Alexendria,[59] a computational database containing over 4.5 million structures, and the ICSD,[61] an experimental database with approximately 300,000 structures. As detailed in Supplementary Table 3, the novelty rate of each method decreased slightly upon comparing to these additional databases. However, the overall trends remain unchanged: random enumeration achieves the highest novelty rate (94.0%) while FTCP exhibits the lowest (35.0%). It is notable that even this lower novelty rate constitutes a substantial fraction of the generated materials, which suggests there remains ample opportunity for materials discovery even as these expansive databases continue to grow. While the novelty rate distribution is interesting, we argue that this is less important than the stability rates since novelty assessments are computationally inexpensive compared to stability assessments, which require DFT calculations.

It is worth noting that the stability rates of novel materials reported in this work are generally lower than the "SUN rates" (stability, uniqueness, and novelty) reported in prior work. For example, MatterGen and CDVAE have previously reported SUN rates of ~38% and ~14%, respectively.[22] Both values are much higher than the 1.8–3.0% stability rates found in our current study. This discrepancy arises primarily from differences in stability criteria. Previous work considered all materials within 100 meV/atom of the convex hull to be "stable." Directly comparable metrics based on this definition are provided in Supplementary Table 4. However, we enforce a stricter definition of stability in the main text of our work, requiring that materials lie on the convex hull ($\Delta E_{\mathrm{d}} \leq 0$) to be considered stable. While many previously synthesized materials are computed to be thermodynamically unstable with DFT, the probability a material can be synthesized is inversely proportional to the magnitude of this instability.[62,63] It is difficult to define a general "rule-of-thumb" for accessible $\Delta E_{\mathrm{d}}$ values, as this will depend on the nature of a material, its competing phases, and available synthetic routes.[43,63,64] Materials computed to be on the hull are likely to be stable at ambient conditions, though synthesizing even hull-stable materials can be challenging.[65–67] We argue that a stricter stability criterion is more meaningful, though a looser cutoff may still be appropriate if one is less concerned with the risk of false positives that would lead to unsuccessful synthesis attempts.

The generative models tested here may not lead to the highest stability rates, but they are unique in their ability to generate new structural frameworks that cannot be mapped to any known prototypes. This sets them apart from baseline methods, which rely entirely on existing templates and therefore exhibit 0% prototype novelty rates (**Table 1**). The generative models evaluated in this work achieve prototype novelty rates ranging from 1.0% (CrystaLLM) to 8.2% (CDVAE). As shown in Supplementary Figure 6, a majority of these materials lie far above the convex hull, exhibiting $\Delta E_{\mathrm{d}}$ > 100 meV/atom. While those from MatterGen are closer to the convex hull on average, none achieve $\Delta E_{\mathrm{d}} \leq 0$. The lack of proposed structures that are both stable and adopt novel prototypes highlights the need for generative models that can effectively balance thermodynamic stability with structural novelty.

## Filtering stability with CHGNet

Filtering materials with uMLIPs such as CHGNet[35] provides a computationally efficient way to improve the stability rate of generation campaigns. Unlike DFT calculations, which are time-consuming and resource-intensive, CHGNet can be used to estimate the internal energy of a material within seconds. This energy can then be compared with a database of DFT-calculated energies to approximate thermodynamic stability. The efficiency of this method allows it to be integrated with any method for generating new materials, whether it be a baseline or generative model. In **Figure 2**, we compare the cumulative distribution functions (CDFs) of DFT-calculated stability results without (left) and with CHGNet filtering (right). In the left panel, materials obtained directly from each method were evaluated by computing their energies with DFT and comparing them to the MP convex hull. In the right panel, only materials predicted to be thermodynamically stable by CHGNet were included for subsequent DFT calculations.

The unfiltered results in the left panel of **Figure 2** serve as a reference to compare each method's ability to generate stable materials. These are CDFs of the same histograms shown in **Figure 1**. As observed in the prior section, ion exchange performs best in generating materials that are stable or close to the convex hull. This is evidenced by a steep rise in its CDF, positioned far to the left of all other methods. There is close competition among the next best three methods – MatterGen, CDVAE, and FTCP – whose CDFs overlap throughout a wide range of energies, reaching 80% near $\Delta E_{\mathrm{d}} \approx 300$ meV/atom. In contrast, CrystaLLM and random enumeration yield CDFs that increase more gradually, reaching 80% only at energies above $\Delta E_{\mathrm{d}} \approx 600$ meV/atom. This suggests that most materials produced by these two methods are unlikely to be accessed experimentally.[63,64]

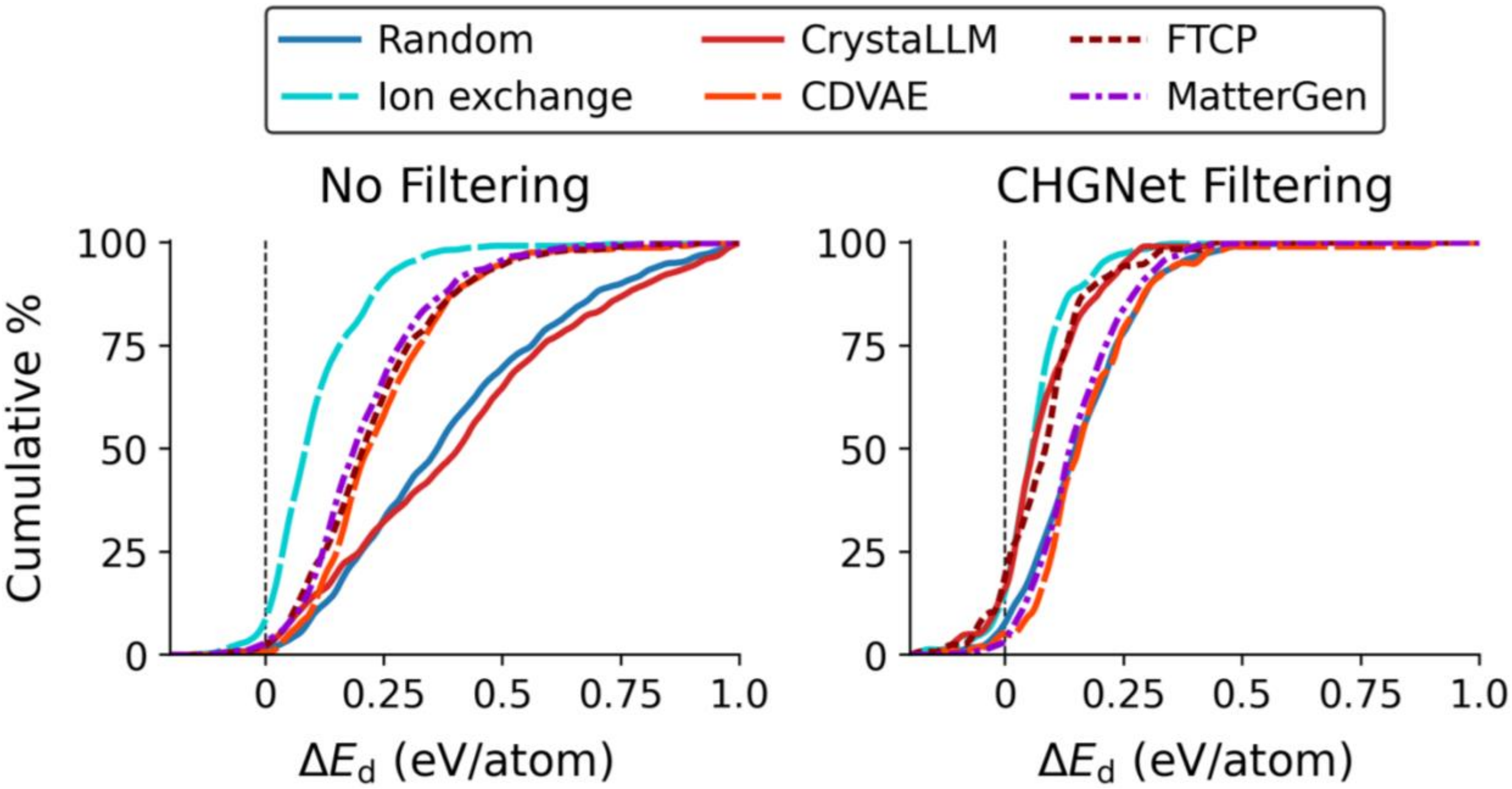

**Figure 2:** Cumulative distribution functions (CDFs) showing the percentage of materials that satisfy a decomposition energy ($\Delta E_\mathrm{d}$) cutoff, with each line color-coded by the method used to generate these materials. The left panel displays CDFs for 500 novel materials generated directly by each method, including two baseline approaches (random enumeration and ion exchange) in blue tones and four generative models (CrystaLLM, CDVAE, FTCP, and MatterGen) in red/purple tones. The right panel displays CDFs for 300 novel materials filtered by CHGNet-predicted stability, including only those CHGNet predicts to have $\Delta E_\mathrm{d} \leq 0$. Filtered energy distributions are also displayed in Supplementary Figure 7.

The right panel of **Figure 2** highlights the beneficial effect of CHGNet filtering, as the consistent leftward shift in all CDFs indicates a greater proportion of materials that are stable or close to the hull. However, results still vary substantially across different generation approaches. Filtered materials from ion exchange show a high stability rate of 15.2%, with the corresponding CDF reaching 80% at $\Delta E_\mathrm{d} \approx 100$ meV/atom. Random enumeration also improves after filtering, achieving a stability rate of 7.6%. Among the generative models, CrystaLLM and FTCP benefit the most from CHGNet filtering, with stability rates increasing to 17.0% and 22.4%, respectively. In contrast, MatterGen and CDVAE show only modest gains, with relatively small shifts in their CDFs and updated stability rates ranging from 3.8% to 8.8%.

We speculate that filtering is less effective for the diffusion models as they often generate materials that fall outside of CHGNet's training distribution – for example, in under-sampled chemistries or structures that are far out-of-equilibrium – potentially reducing the accuracy of stability predictions and limiting their performance gains. Indeed, Supplementary Figure 8 shows large mean absolute errors (154 to 156 meV/atom) on structures from CDVAE and MatterGen. Large prediction errors (139 meV/atom) are also observed on structures generated through random enumeration. Despite adhering to known structure templates, random enumeration more often produces exotic compositions with less representation in CHGNet's training set.

To assess the compositional diversity of generated materials, we provide heatmaps of element frequencies and histograms showing the number of elements per novel compound in Supplementary Figures 9-10. Random enumeration produces compositions spanning much of the periodic table with relatively even distributions of pnictides, chalcogenides, and halides. However, these compositions are generally limited to ternary prototypes, reflecting the dominance of known three-element structures. In contrast, ion exchange produces a more diverse set of compounds – including quaternaries and quinaries – but the overall composition space is narrower, skewed toward oxides which are disproportionately prevalent in MP. This contributes to ion exchange providing the lowest CHGNet prediction error (47 meV/atom) among all methods.

Similarly, most of the generative models produce oxides at disproportionately high rates – reflecting bias in the MP-20 dataset on which they were all trained. For example, CDVAE generates a relatively narrow range of compositions, with 23.4% containing oxygen. However, it also produces more complex chemical formulae than template-based methods, with up to nine elements per compound. These multicomponent oxides are more often novel than compositions with fewer elements; however, they also compete with many compounds in the high-dimensional phase diagram, which contributes to the lower stability rate of CDVAE. Other methods generate a balance of binaries, ternaries, and quaternaries with broad periodic table coverage but a slight preference for oxides and halides. This further reflects bias in the MP-20 dataset used for training, which can be mitigated by expanding the set to include more diverse compounds. For instance, the proportion of oxides generated by MatterGen drops from 21.2% to 10.9% after incorporating the Alexandria dataset into its training. FTCP demonstrates the most compositional diversity of the models tested here, sampling a wide range of elements with only 4.4% of its materials containing oxygen. Although FTCP is trained on the MP-20 dataset, we suspect its strategy of latent space sampling enables it to interpolate between known compounds and explore regions of composition space that are underrepresented in the training data.

## Generating materials with targeted properties

Beyond generating thermodynamically stable materials, it is also useful to generate materials with targeted properties for particular applications. To this end, we adapted our baseline methods and re-trained a generative model to target desired band gaps and bulk moduli, aligning with the property-driven generation strategies demonstrated in recent work.[22] First targeting materials with a band gap near 3 eV, two baseline methods were applied: random enumeration and ion exchange. Additionally, we applied ML filtering (based on predictions from CGCNN)[36] to the randomly enumerated materials to assess its impact on targeting specific properties. Finally, we tested one generative model, FTCP, whose latent space can be conditioned on specific properties. A total of 500 novel materials were generated from each method, and their distributions of computed band gaps are shown in **Figure 3**.

Random enumeration produced a wide variety of materials, with 30.2% of the novel ones being metallic. Only 11.2% of these materials exhibited a band gap within 0.5 eV of the desired value (3 eV), demonstrating the low success rate of computational screening when no guidance is provided. Applying CGCNN to filter these randomly enumerated materials improved the results considerably. By only retaining materials with CGCNN-predicted band gaps near 3 eV, the proportion of metals dropped to 18.8%, and 21.4% of the filtered materials exhibited band gaps within 0.5 eV of the target. As with CHGNet-filtering, this showcases the utility of ML-based screening for quickly refining large pools of candidate materials.

Data-driven ion exchange performed even better than CGCNN filtering of randomly enumerated materials, leveraging its ability to generate hypothetical compounds by substituting ions in known materials from MP that already have band gaps close to 3 eV. This method resulted in only 5.6% of the novel materials being metallic and a substantial 37.2% of them having a band gap within 0.5 eV of the target. This strong performance may not be entirely surprising as many of the compositional changes introduced by ion exchange are relatively minor, especially when the substituted element constitutes a small fraction of the overall chemical formula. This mirrors our findings from the previous section, highlighting the tradeoff between achieving success – whether in targeted properties or stability – and prioritizing novelty or diversity in the generated structures.

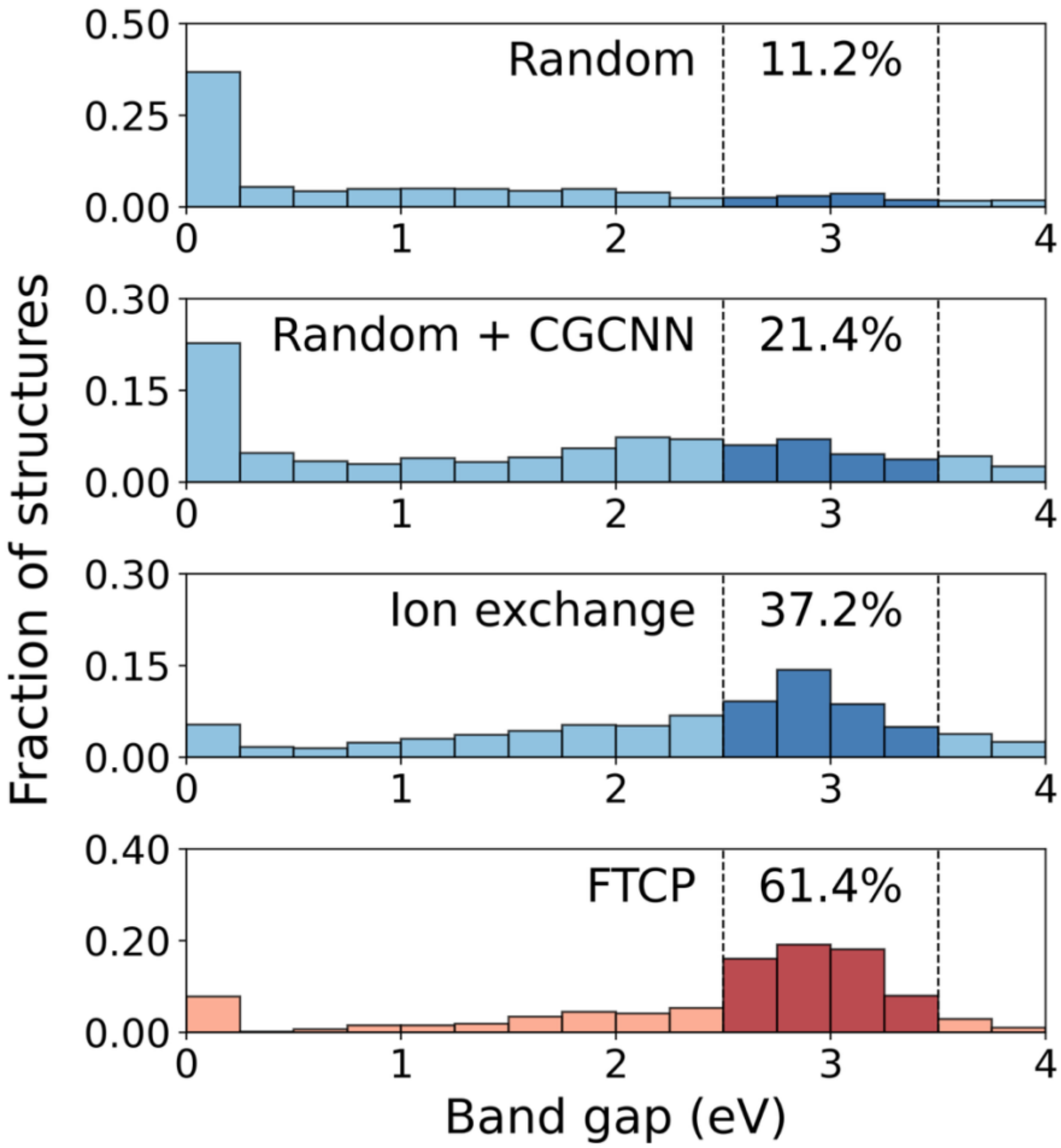

**Figure 3:** Histograms showing density functional theory (DFT) computed band gap distributions of structures generated by two baseline methods (random enumeration and ion exchange, colored blue), CGCNN (also colored blue) applied to filter the randomly enumerated materials, and one generative model: FTCP (colored red). For each of the four approaches, 500 novel materials were considered. With the exception of random enumeration, all methods specifically targeted materials with a band gap near 3 eV. The percentage of generated materials with a band gap in the range of 2.5 to 3.5 eV is displayed above the shaded bars in each subplot.

FTCP outperformed all other methods in targeting electronic band gap, with 61.4% of its novel materials exhibiting a band gap within 0.5 eV of the desired value (3 eV). This success likely stems from FTCP's latent space sampling informed by known compounds with band gaps close to the target, which enables the generation of materials with structural or compositional similarities to the reference points. Thermodynamic stability remains an important consideration, as only 3.0% of the novel materials generated by FTCP are stable, compared with 15.2% of those generated by ion exchange ($\Delta E_{\mathrm{d}}$ distributions provided in Supplementary Figure 11).

Using the same four methods described above (for targeting a desired band gap), we next generated materials with the objective of maximizing bulk modulus. This task fundamentally differs from the previous band gap-related objective by focusing on materials with extreme properties (*e.g.*, maximal bulk modulus) instead of those within an intermediate range (*e.g.*, band gaps near 3 eV). A total of 500 materials were sampled from each method, and their bulk moduli were computed using Birch-Murnaghan equations of state fit to DFT-computed energies. The resulting distributions of bulk moduli are shown in **Figure 4**. Materials generated through random enumeration follow a Poisson-like distribution of bulk moduli with a peak near 50–60 GPa, closely resembling the known distribution of elastic properties for materials in MP.[68] If we define success as finding novel materials with a bulk modulus ≥ 300 GPa, then random enumeration achieves this at a rate of only 3.0%. When CGCNN is applied to filter these materials, it causes a noticeable shift in the distribution toward higher bulk moduli, and 15.4% of the filtered materials exhibit a bulk modulus ≥ 300 GPa.

When applied to known materials in MP with high bulk moduli, ion exchange performs more modestly, with 8.6% of the resulting materials exhibiting a bulk modulus ≥ 300 GPa. This smaller shift in the distribution likely reflects the tendency for ion exchange to introduce only minor compositional changes, which limits its ability to substantially alter the mechanical properties of the original materials – many of which (in MP) do not exhibit anomalously high bulk moduli. FTCP performed slightly better than ion exchange but worse than CGCNN-based filtering of randomly enumerated materials, with 9.2% of the compounds generated by FTCP exhibiting a bulk modulus ≥ 300 GPa.

Compared to its strong performance on electronic band gap, we suspect FTCP is less effective here given the scarcity of materials with extremely high bulk moduli in MP. This lack of training data may limit the conditioning of the autoencoder's latent space on extreme bulk modulus values. FTCP also yields a low stability rate of 2.0% when targeting novel materials with high bulk modulus. CGCNN filtering and ion exchange face similar limitations, with stability rates of 2.0% and 1.8%, respectively ($\Delta E_{\mathrm{d}}$ distributions provided in Supplementary Figure 12). These uniformly low percentages across all evaluated methods highlight the challenge of identifying

"exceptional" materials, as the inherent scarcity of analogs in the materials space and limited training data inhibit the development of effective models for both generation and filtering.[69]

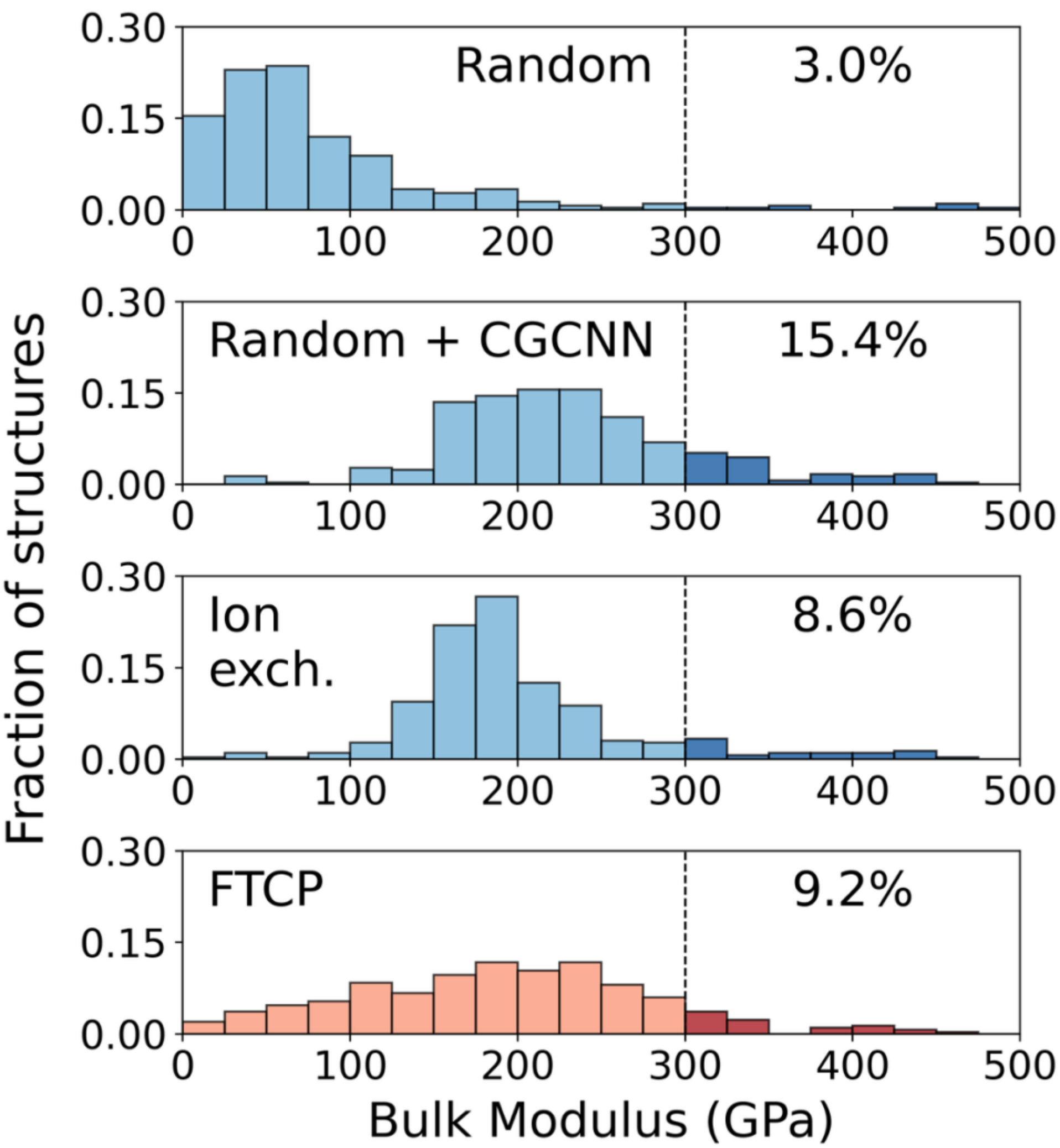

**Figure 4:** Histograms showing density functional theory (DFT) computed bulk moduli distributions of structures generated by two baseline methods (random enumeration and ion exchange, colored blue), CGCNN (also colored blue) applied to filter the randomly enumerated materials, and one generative model: FTCP (colored red). For each of the four approaches, 500 novel materials were considered. With the exception of random enumeration, all methods specifically targeted materials with a bulk modulus ≥ 300 GPa. The percentage of generated materials satisfying this objective is displayed above the shaded bars in each subplot.

## Discussion

The recent surge in generative models for inorganic crystalline materials underscores the growing need for benchmarks to assess their performance. While metrics such as stability and novelty provide valuable insight, there is a lack of clear baselines for comparison. To address this gap, we developed and evaluated two baseline methods: random enumeration of charge-balanced compositions in known structure prototypes, and targeted ion substitution of known materials with desired properties. These approaches leverage existing data from AFLOW[33,34] and the Materials Project,[29] which offer a wealth of information on structure prototypes and calculated properties acquired from DFT calculations. They also benefit from simple yet powerful chemical heuristics; charge balance favors validity of proposed chemical formulae, and substitutions are performed on ions of comparable size and oxidation state. As a result, the baseline methods perform well in generating stable materials not found in existing databases. Random enumeration achieves a modest stability rate of 1.4%, while ion exchange achieves a much higher rate of 9.2%. There remains ample opportunity to further increase these rates as additional chemical heuristics are introduced to better filter computationally proposed materials.[70]

The strong performance of the baseline methods establishes a high benchmark for generative models to meet or exceed. For this task, we tested a variational autoencoder (FTCP),[12] a large-language model (CrystaLLM),[15] and two diffusion models (CDVAE and MatterGen).[13,22] Our tests showed MatterGen to be most effective in generating materials on or close to the hull, though its stability rate of 3.0% still falls well below that of ion exchange (9.2%). Nevertheless, generative models excel in generating materials with a high degree of structural novelty; up to 8.2% cannot be mapped to any known structure prototype in the AFLOW database. The capability of generating entirely new structural arrangements is unique to the generative models, but their low stability rates leave much room for improvement. One promising direction is to expand the training data for these models. For example, we found that MatterGen achieves a higher stability rate of 5.4% when trained on materials from Alexandria in addition to MP-20. Similar improvements were observed for CrystaLLM.

It is important to note that many of the comparisons made in this work depend on the stability threshold used to define success. We adopted a strict criterion requiring materials to lie on the convex hull ($\Delta E_{\mathrm{d}} \leq 0$) to be considered stable, which reduces false positives but also penalizes the generation of near-stable candidates that may be synthesizable. A looser threshold would raise the stability rates across all methods and shift the relative performance of each approach. For example, a threshold of $\Delta E_{\mathrm{d}} \leq 100$ meV/atom increases the stability rate of the generative models to 11-20%, but these rates still far well below ion exchange, which achieves a stability rate of 58% using the same threshold (Supplementary Table 4).

In addition to generating a large proportion of materials near the convex hull, generative models also perform well in targeting specific properties when sufficient training data is available. For instance, FTCP achieves a high success rate of 61.4% in generating materials with a desired band gap near 3 eV, far surpassing the 37.2% achieved by ion exchange. This performance diminishes when targeting extreme values of properties such as high bulk moduli (> 300 GPa) that are less well represented in the training set. However, improved results can likely be obtained by running additional calculations on materials with extreme property and feeding them back into the generative models as training data.

To enhance the performance of the methods discussed in this paper, machine learning models were used to filter proposed materials based on predicted thermodynamic stability or desired properties. Our results demonstrate that this is a highly effective approach. Filtering by predicted stability using a pre-trained uMLIP (CHGNet)[35] substantially improves the stability rates of generated materials. A notably high 22.4% of novel materials generated by FTCP lie on the DFT convex hull after filtering. This performance boost achieved by filtering is diminished for some generative models like CDVAE and MatterGen, which produce more exotic materials that fall outside of CHGNet's training distribution. However, uMLIPs are likely to become more effective at filtering such materials as the breadth and diversity of their training data improves.[71] This trend is evident in the correlation between prediction error and the filtered stability rate (Supplementary Figure 13), suggesting that reducing prediction error in future uMLIPs should further improve stability rates.

Similar findings were observed when using a pre-trained graph neural network (CGCNN)[36] to filter materials by predicted band gap and bulk modulus. Doing so leads to a near three-fold increase in the success rate of identifying materials with desired properties compared to random enumeration but remains relatively low (15.4%) when targeting extreme property values (*e.g.*, a bulk modulus > 300 GPa). It also leads to a decrease in the stability rate of the proposed materials, though incorporating a uMLIP-based stability filter could mitigate this issue. As with uMLIPs, these findings underscore the need to broaden and diversify training data for property prediction models to enhance the efficiency of generative approaches in identifying novel materials with exceptional properties.

Our findings demonstrate that there is still room for improvement in the design of generative models for inorganic materials, particularly when they are used to find new materials that are thermodynamically stable. To streamline the development of future models, we provide all of the data and code from this work in a publicly accessible GitHub repository (see Data Availability Statement). We envision these resources being used for benchmarking generative models and integrating them with traditional screening methods to enhance the success rate in discovering new materials that are likely to be synthesized and display desired properties.

**Data Availability Statement**

The code for generating materials through random enumeration and ion exchange, in addition to filtering these materials on the basis of machine learning predictions, is available at https://github.com/Bartel-Group/matgen_baselines. This repository also includes CIF files for the structures generated by each method, as well as pre-trained models on the MP-20 dataset and installation instructions for CrystaLLM, FTCP, CDVAE, and MatterGen.

**Acknowledgements**

This work was supported by the 3DEAP NRT, NSF grant no. 2345719, and new faculty start-up funds from the University of Minnesota. The authors also acknowledge the Minnesota Supercomputing Institute (MSI) at the University of Minnesota for providing resources that contributed to the research results reported herein. This work was also enabled by the dedication and open-source contributions of prior developers, including those behind the pre-trained graph neural networks (CHGNet and CGCNN), generative models (CrystaLLM, CDVAE, FTCP, and MatterGen), structure prediction algorithms (pymatgen), structure prototypes (AFLOW), and DFT databases (Materials Project).

**References**

1. Cheetham, A. K., Seshadri, R. & Wudl, F. Chemical synthesis and materials discovery. *Nat. Synth* **1**, 514–520 (2022).
2. A. Manthiram, J. C. Knight, S.-T. Myung, S.-M. Oh, & Y-K. Sun. Nickel-Rich and Lithium-Rich Layered Oxide Cathodes: Progress and Perspectives. *Adv. Energy Mater.* **6**, 1501010 (2016).

3. K. Nomura *et al.* Thin-Film Transistor Fabricated in Single-Crystalline Transparent Oxide Semiconductor. *Science* **300**, 1269–1272 (2003).

4. K. M. Shen & J. C. Seamus Davis. Cuprate high-Tc superconductors. *Mater. Today* **11**, 14–21 (2008).

5. Park, H., Li, Z. & Walsh, A. Has generative artificial intelligence solved inverse materials design? *Matter* **7**, 2355–2367 (2024).

6. Wang, Z., Hua, H., Lin, W., Yang, M. & Tan, K. C. Crystalline Material Discovery in the Era of Artificial Intelligence. Preprint at https://doi.org/10.48550/arXiv.2408.08044 (2025).

7. A. Nouira, N. Sokolovska, & J.-C. Crivello. CrystalGAN: Learning to Discover Crystallographic Structures with Generative Adversarial Networks. *arXiv:1810.11203* (2018).

8. S. Kim, J. Noh, G. Gu, A. Aspuru-Guzik, & Y. Jung. Generative Adversarial Networks for Crystal Structure Prediction. *ACS Cent. Sci.* **6**, 1412–1420 (2020).

9. Y. Zhao *et al.* High-Throughput Discovery of Novel Cubic Crystal Materials Using Deep Generative Neural Networks. *Adv. Sci.* **8**, 2100566 (2021).

10. J. Noh *et al.* Inverse Design of Solid-State Materials via a Continuous Representation. *Matter* **1**, 1370–1384 (2019).

11. Court, C. J., Yildirim, B., Jain, A. & Cole, J. M. 3-D Inorganic Crystal Structure Generation and Property Prediction via Representation Learning. *J. Chem. Inf. Model.* **60**, 4518–4535 (2020).

12. Ren, Z. *et al.* An invertible crystallographic representation for general inverse design of inorganic crystals with targeted properties. *Matter* **5**, 314–335 (2022).

13. Xie, T., Fu, X., Ganea, O.-E., Barzilay, R. & Jaakkola, T. Crystal Diffusion Variational Autoencoder for Periodic Material Generation. Preprint at https://doi.org/10.48550/arXiv.2110.06197 (2022).

14. Zhu, R., Nong, W., Yamazaki, S. & Hippalgaonkar, K. WyCryst: Wyckoff inorganic crystal generator framework. *Matter* **7**, 3469–3488 (2024).

15. Antunes, L. M., Butler, K. T. & Grau-Crespo, R. Crystal Structure Generation with Autoregressive Large Language Modeling. Preprint at https://doi.org/10.48550/arXiv.2307.04340 (2024).

16. Flam-Shepherd, D. & Aspuru-Guzik, A. Language models can generate molecules, materials, and protein binding sites directly in three dimensions as XYZ, CIF, and PDB files. Preprint at https://doi.org/10.48550/arXiv.2305.05708 (2023).

17. Gruver, N. *et al.* Fine-Tuned Language Models Generate Stable Inorganic Materials as Text. Preprint at https://doi.org/10.48550/arXiv.2402.04379 (2024).

18. Breuck, P.-P. D., Piracha, H. A., Rignanese, G.-M. & Marques, M. A. L. A generative material transformer using Wyckoff representation. Preprint at https://doi.org/10.48550/arXiv.2501.16051 (2025).

19. Mohanty, T., Mehta, M., Sayeed, H. M., Srikumar, V. & Sparks, T. D. CrysText: A Generative AI Approach for Text-Conditioned Crystal Structure Generation using LLM. Preprint at https://doi.org/10.26434/chemrxiv-2024-gjhpq (2024).

20. Alverson, M. *et al.* Generative adversarial networks and diffusion models in material discovery. *Digital Discovery* **3**, 62–80 (2024).

21. Jiao, R. *et al.* Crystal Structure Prediction by Joint Equivariant Diffusion. Preprint at https://doi.org/10.48550/arXiv.2309.04475 (2024).

22. Zeni, C. *et al.* A generative model for inorganic materials design. *Nature* **639**, 624–632 (2025).

23. Yang, S. *et al.* Scalable Diffusion for Materials Generation. Preprint at https://doi.org/10.48550/arXiv.2311.09235 (2024).

24. Hoellmer, P. *et al.* Open Materials Generation with Stochastic Interpolants. Preprint at https://doi.org/10.48550/arXiv.2502.02582 (2025).

25. Miller, B. K., Chen, R. T. Q., Sriram, A. & Wood, B. M. FlowMM: Generating Materials with Riemannian Flow Matching. Preprint at https://doi.org/10.48550/arXiv.2406.04713 (2024).

26. Sriram, A., Miller, B. K., Chen, R. T. Q. & Wood, B. M. FlowLLM: Flow Matching for Material Generation with Large Language Models as Base Distributions. Preprint at https://doi.org/10.48550/arXiv.2410.23405 (2024).

27. Ruple, L., Torresi, L., Schopmans, H. & Friederich, P. Symmetry-Aware Bayesian Flow Networks for Crystal Generation. Preprint at https://doi.org/10.48550/arXiv.2502.03146 (2025).

28. Tangsongcharoen, K. *et al.* CrystalGRW: Generative Modeling of Crystal Structures with Targeted Properties via Geodesic Random Walks. Preprint at https://doi.org/10.48550/arXiv.2501.08998 (2025).

29. A. Jain *et al.* Commentary: The Materials Project: A materials genome approach to accelerating materials innovation. *APL Mater.* **1**, 011002 (2013).

30. Park, H., Onwuli, A. & Walsh, A. Exploration of crystal chemical space using text-guided generative artificial intelligence. Preprint at https://doi.org/10.26434/chemrxiv-2024-rw8p5 (2024).

31. Baird, S. G., Sayeed, H. M., Montoya, J. & Sparks, T. D. matbench-genmetrics: A Python library for benchmarking crystal structure generative models using time-based splits of Materials Project structures. *Journal of Open Source Software* **9**, 5618 (2024).

32. Riebesell, J. *et al.* Matbench Discovery -- A framework to evaluate machine learning crystal stability predictions. Preprint at https://doi.org/10.48550/arXiv.2308.14920 (2024).

33. Mehl, M. J. *et al.* The AFLOW Library of Crystallographic Prototypes: Part 1. *Computational Materials Science* **136**, S1–S828 (2017).

34. Hicks, D. *et al.* The AFLOW Library of Crystallographic Prototypes: Part 2. *Computational Materials Science* **161**, S1–S1011 (2019).

35. B. Deng *et al.* CHGNet as a pretrained universal neural network potential for charge-informed atomistic modelling. *Nat. Mach. Intell.* **5**, 1031–1041 (2023).

36. T. Xie & J. C. Grossman. Crystal Graph Convolutional Neural Networks for an Accurate and Interpretable Prediction of Material Properties. *Phys. Rev. Lett.* **120**, 145301 (2018).

37. S. P. Ong *et al.* Python Materials Genomics (pymatgen): A robust, open-source python library for materials analysis. *Comput. Mater. Sci.* **68**, 314–319 (2013).

38. Hautier, G., Fischer, C., Ehrlacher, V., Jain, A. & Ceder, G. Data Mined Ionic Substitutions for the Discovery of New Compounds. *Inorg. Chem.* **50**, 656–663 (2011).

39. Levin, I. NIST Inorganic Crystal Structure Database (ICSD). National Institute of Standards and Technology https://doi.org/10.18434/M32147 (2020).

40. Perdew, J. P., Burke, K. & Ernzerhof, M. Generalized Gradient Approximation Made Simple. *Phys. Rev. Lett.* **77**, 3865–3868 (1996).

41. Kresse, G. & Furthmüller, J. Efficiency of ab-initio total energy calculations for metals and semiconductors using a plane-wave basis set. *Computational Materials Science* **6**, 15–50 (1996).

42. Kresse, G. & Furthmüller, J. Efficient iterative schemes for ab initio total-energy calculations using a plane-wave basis set. *Phys. Rev. B* **54**, 11169–11186 (1996).

43. C. J. Bartel. Review of computational approaches to predict the thermodynamic stability of inorganic solids. *J. Mater. Sci.* **57**, 10475–10498 (2022).

44. Wang, A. *et al.* A framework for quantifying uncertainty in DFT energy corrections. *Sci Rep* **11**, 15496 (2021).

45. Birch, F. Finite Elastic Strain of Cubic Crystals. *Phys. Rev.* **71**, 809–824 (1947).

46. Jacobs, R. *et al.* A practical guide to machine learning interatomic potentials – Status and future. *Current Opinion in Solid State and Materials Science* **35**, 101214 (2025).

47. Larsen, A. H. *et al.* The atomic simulation environment—a Python library for working with atoms. *J. Phys.: Condens. Matter* **29**, 273002 (2017).

48. Zhang, X., Stevanović, V., d'Avezac, M., Lany, S. & Zunger, A. Prediction of ${A}_{2}B{X}_{4}$ metal-chalcogenide compounds via first-principles thermodynamics. *Phys. Rev. B* **86**, 014109 (2012).

49. Davies, D. W. *et al.* Computational Screening of All Stoichiometric Inorganic Materials. *Chem* **1**, 617–627 (2016).

50. Gorai, P., Ganose, A., Faghaninia, A., Jain, A. & Stevanović, V. Computational discovery of promising new n-type dopable ABX Zintl thermoelectric materials. *Mater. Horiz.* **7**, 1809–1818 (2020).

51. Wang, H.-C., Botti, S. & Marques, M. A. L. Predicting stable crystalline compounds using chemical similarity. *npj Comput Mater* **7**, 12 (2021).

52. Vasylenko, A. *et al.* Element selection for crystalline inorganic solid discovery guided by unsupervised machine learning of experimentally explored chemistry. *Nat Commun* **12**, 5561 (2021).

53. Merchant, A. *et al.* Scaling deep learning for materials discovery. *Nature* **624**, 80–85 (2023).

54. Emery, A. A., Saal, J. E., Kirklin, S., Hegde, V. I. & Wolverton, C. High-Throughput Computational Screening of Perovskites for Thermochemical Water Splitting Applications. *Chem. Mater.* **28**, 5621–5634 (2016).

55. Wang, Y., Baldassarri, B., Shen, J., He, J. & Wolverton, C. Landscape of Thermodynamic Stabilities of A2BB′O6 Compounds. *Chem. Mater.* **36**, 6816–6830 (2024).

56. Schmidt, J. *et al.* Predicting the Thermodynamic Stability of Solids Combining Density Functional Theory and Machine Learning. *Chem. Mater.* **29**, 5090–5103 (2017).

57. Kocevski, V., Pilania, G. & Uberuaga, B. P. High-throughput investigation of the formation of double spinels. *J. Mater. Chem. A* **8**, 25756–25767 (2020).

58. Shi, J. *et al.* High-throughput search of ternary chalcogenides for p-type transparent electrodes. *Sci Rep* **7**, 43179 (2017).

59. Schmidt, J. *et al.* Improving machine-learning models in materials science through large datasets. *Materials Today Physics* **48**, 101560 (2024).

60. T. Butler, K., M. Frost, J., M. Skelton, J., L. Svane, K. & Walsh, A. Computational materials design of crystalline solids. *Chemical Society Reviews* **45**, 6138–6146 (2016).

61. Zagorac, D., Müller, H., Ruehl, S., Zagorac, J. & Rehme, S. Recent developments in the Inorganic Crystal Structure Database: theoretical crystal structure data and related features. *J Appl Cryst* **52**, 918–925 (2019).

62. Narayan, A. *et al.* Computational and experimental investigation for new transition metal selenides and sulfides: The importance of experimental verification for stability. *Phys. Rev. B* **94**, 045105 (2016).

63. W. Sun *et al.* The thermodynamic scale of inorganic crystalline metastability. *Sci. Adv.* **2**, e160022 (2016).

64. M. Aykol, S. S. Dwaraknath, W. Sun, & K. A. Persson. Thermodynamic limit for synthesis of metastable inorganic materials. *Sci. Adv.* **4**, eaaq014 (2018).

65. Wustrow, A. *et al.* Synthesis and Characterization of MgCr2S4 Thiospinel as a Potential Magnesium Cathode. *Inorg. Chem.* **57**, 8634–8638 (2018).

66. A. Miura *et al.* Selective metathesis synthesis of MgCr2S4 by control of thermodynamic driving forces. *Mater. Horiz.* **7**, 1310–1316 (2020).

67. Pandey, S. *et al.* Steam-Assisted Ammonolysis of MoO2 as a Synthetic Pathway to Oxygenated δ-MoN. *Materials* **18**, 2340 (2025).

68. de Jong, M. *et al.* Charting the complete elastic properties of inorganic crystalline compounds. *Sci Data* **2**, 150009 (2015).

69. Schrier, J., Norquist, A. J., Buonassisi, T. & Brgoch, J. In Pursuit of the Exceptional: Research Directions for Machine Learning in Chemical and Materials Science. *J. Am. Chem. Soc.* **145**, 21699–21716 (2023).

70. Das, B., Ji, K., Sheng, F., McCall, K. M. & Buonassisi, T. Embedding human knowledge in material screening pipeline as filters to identify novel synthesizable inorganic materials. *Faraday Discuss.* (2024) doi:10.1039/D4FD00120F.

71. Barroso-Luque, L. *et al.* Open Materials 2024 (OMat24) Inorganic Materials Dataset and Models. Preprint at https://doi.org/10.48550/arXiv.2410.12771 (2024).

# Supplementary Information

# Establishing baselines for generative discovery of inorganic crystals

Nathan J. Szymanski[1] and Christopher J. Bartel[1,*]

[1] University of Minnesota, Department of Chemical Engineering and Materials Science, Minneapolis, MN, USA 55455

[*] Correspondence to cbartel@umn.edu

**Supplementary Table 1:** Average times required by a MacBook Air (Apple M2 chip, 16 GB RAM) to generate a single material using each method. These runs were performed sequentially on a single core of the M2 chip. MatterGen is excluded from this table as we ran it on a single GPU (NVIDIA A100), requiring ~9.22 seconds per material generated. Otherwise, it was much too slow (several minutes per material) to run in high throughput.

| **Method** | **Time per structure (s)** |
|---|---|
| Random | 0.14 |
| Ion exchange | 3.45 |
| CrystaLLM | 38.27 |
| CDVAE | 14.32 |
| FTCP | 0.03 |

**Supplementary Table 2:** Stability and novelty rates of materials generated by FTCP as a function of sampling radius, which determines the distance from known materials in the latent space for sampling before decoding *via* the variational autoencoder. These statistics are based on 500 materials when using a sampling radius of 0.6, and 100 materials when using larger sampling radii (1.2, 1.8, and 2.4). Only the novel materials are considered when computing stability rate.

| Sampling radius | Stability rate | Novelty rate |
|---|---|---|
| 0.6 | 2.0% | 38.2% |
| 1.2 | 1.0% | 72.0% |
| 1.8 | 0.0% | 91.0% |
| 2.4 | 0.0% | 95.0% |

**Supplementary Table 3:** Each novelty rate describes the percentage of materials generated by each method that are absent from the Materials Project, Alexandria, and the Inorganic Crystal Structure Database (ICSD). The combined novelty rate accounts for absence from all three datasets. Structural similarity was assessed using the *StructuerMatcher* tool in *pymatgen*. We did not include any comparisons with disordered materials.

| Method | Materials Project novelty rate | Alexandria novelty rate | ICSD novelty rate | Combined novelty rate |
|---|---|---|---|---|
| Random | 98.6% | 95.2% | 98.8% | 94.0% |
| Ion exchange | 72.4% | 69.6% | 78.4% | 62.8% |
| CrystaLLM | 98.2% | 90.8% | 95.4% | 85.0% |
| CDVAE | 96.0% | 96.0% | 96.6% | 92.0% |
| FTCP | 38.2% | 40.2% | 61.8% | 35.0% |
| MatterGen | 91.8% | 86.4% | 92.2% | 83.2% |

**Supplementary Table 4:** Percentage of materials generated by each method that are within 100 meV/atom of the convex hull defined by entries from the Materials Project. This corresponds to the "stability rate" following common definitions in past work (*e.g.*, CDVAE and MatterGen).

| Method | Materials with $\Delta E_d \leq 100$ meV/atom |
|---|---|
| Random | 8.2% |
| Ion exchange | 58.4% |
| CrystaLLM | 12.8% |
| CDVAE | 11.2% |
| FTCP | 20.0% |
| MatterGen | 17.8% |

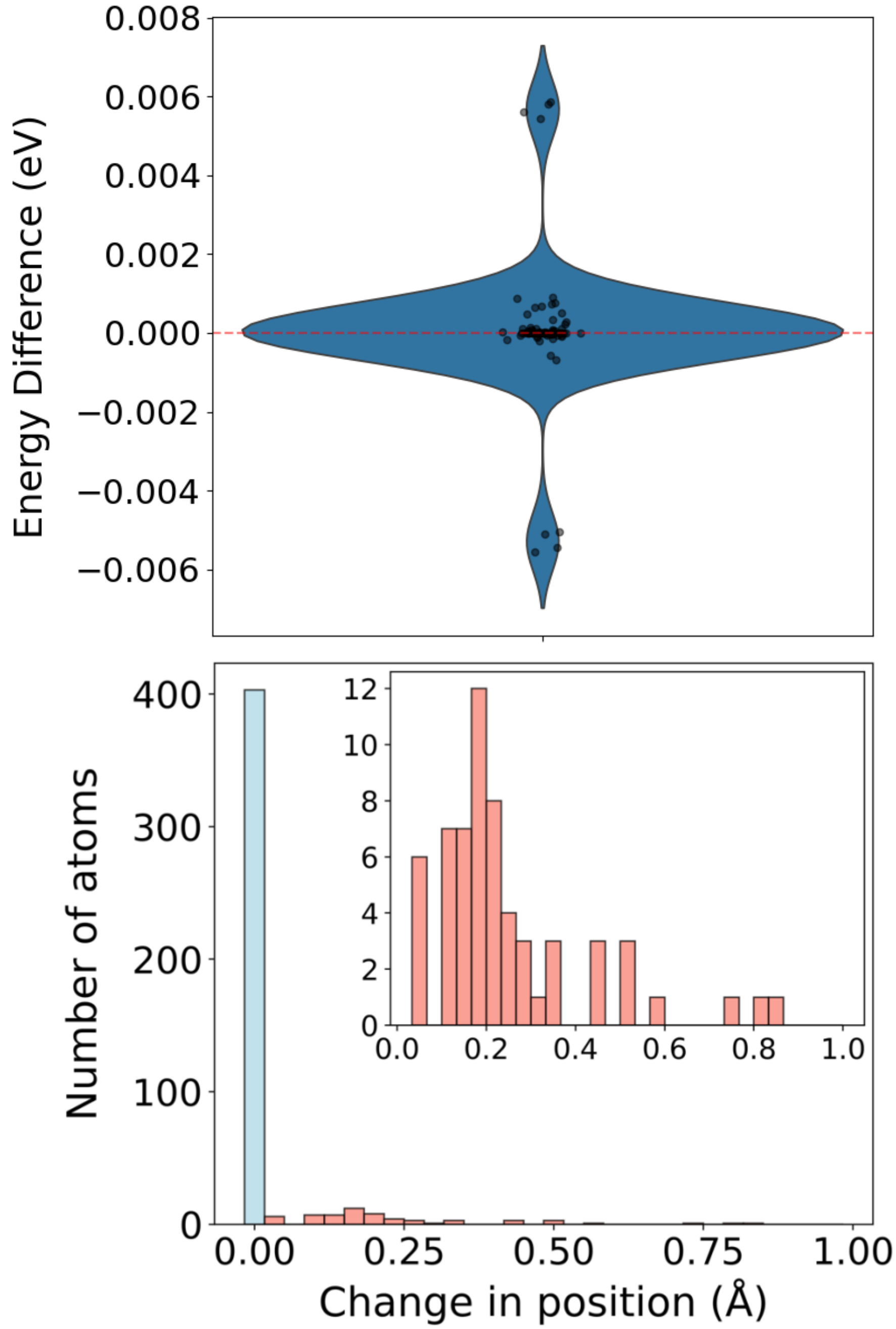

**Supplementary Figure 1:** (Top) Violin and scatter plot showing total energy differences between 80 structures – spanning 52 elements and 21 space groups (from $P\bar{1}$ to $Fm\bar{3}m$) – relaxed using a loose (0.03 eV/Å) versus tight (0.001 eV/Å) convergence criterion. About 90% of the structures

exhibit a change in energy < 0.001 eV when a tighter convergence criterion was used. Larger changes are observed for eight structures but remain < 0.007 eV. (Bottom) Changes in the atomic positions of these same 80 structures, relaxed using loose versus tight convergence criterion, are shown as a histogram. The inset (red bars) shows changes with magnitude larger than 0.01 Å.

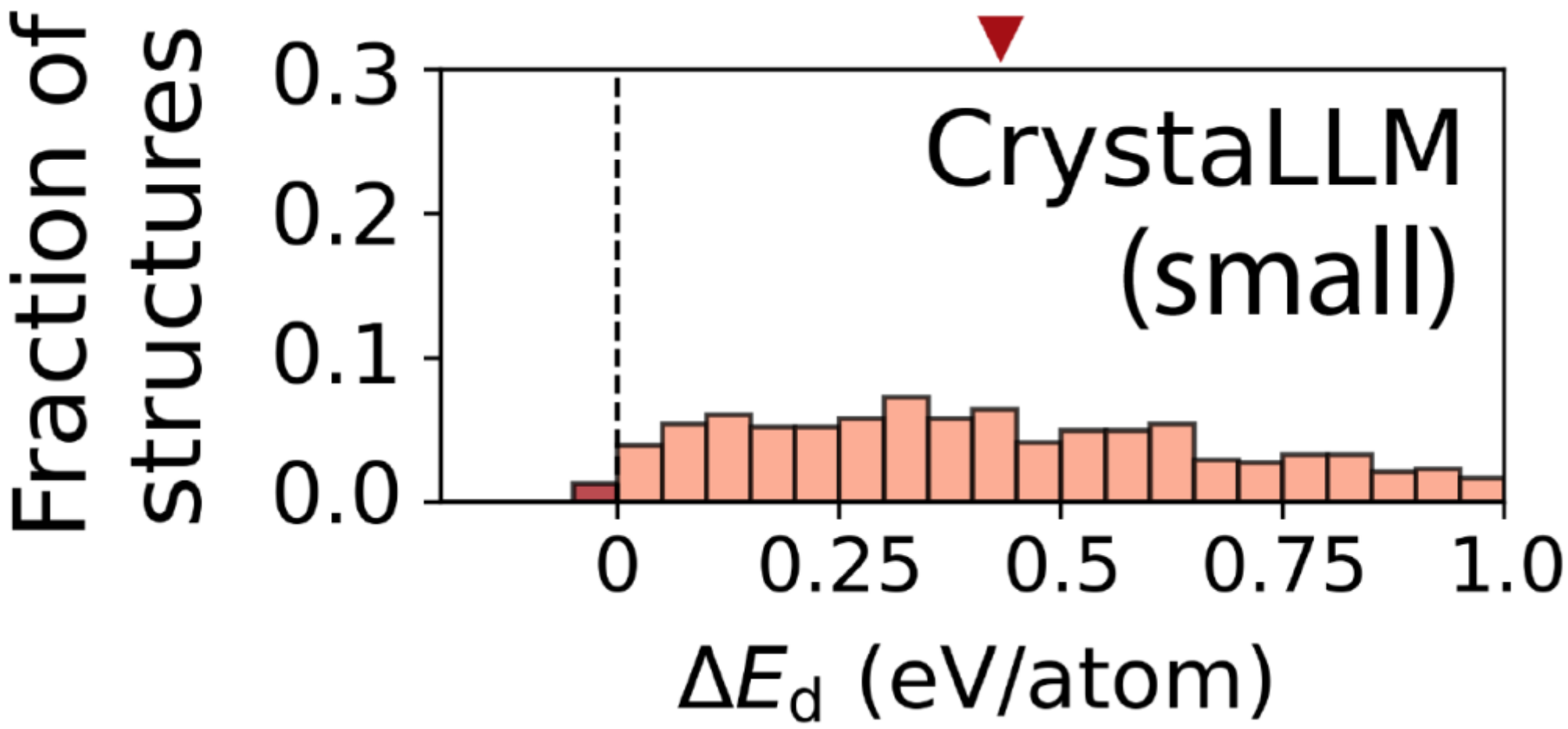

**Supplementary Figure 2:** Histograms showing density functional theory (DFT) computed decomposition energies of novel structures generated using a small version (25 million parameters) of CrystaLLM (from https://github.com/lantunes/CrystaLLM) trained on the MP-20 dataset. The triangular marker indicates the median decomposition energy of generated structures.

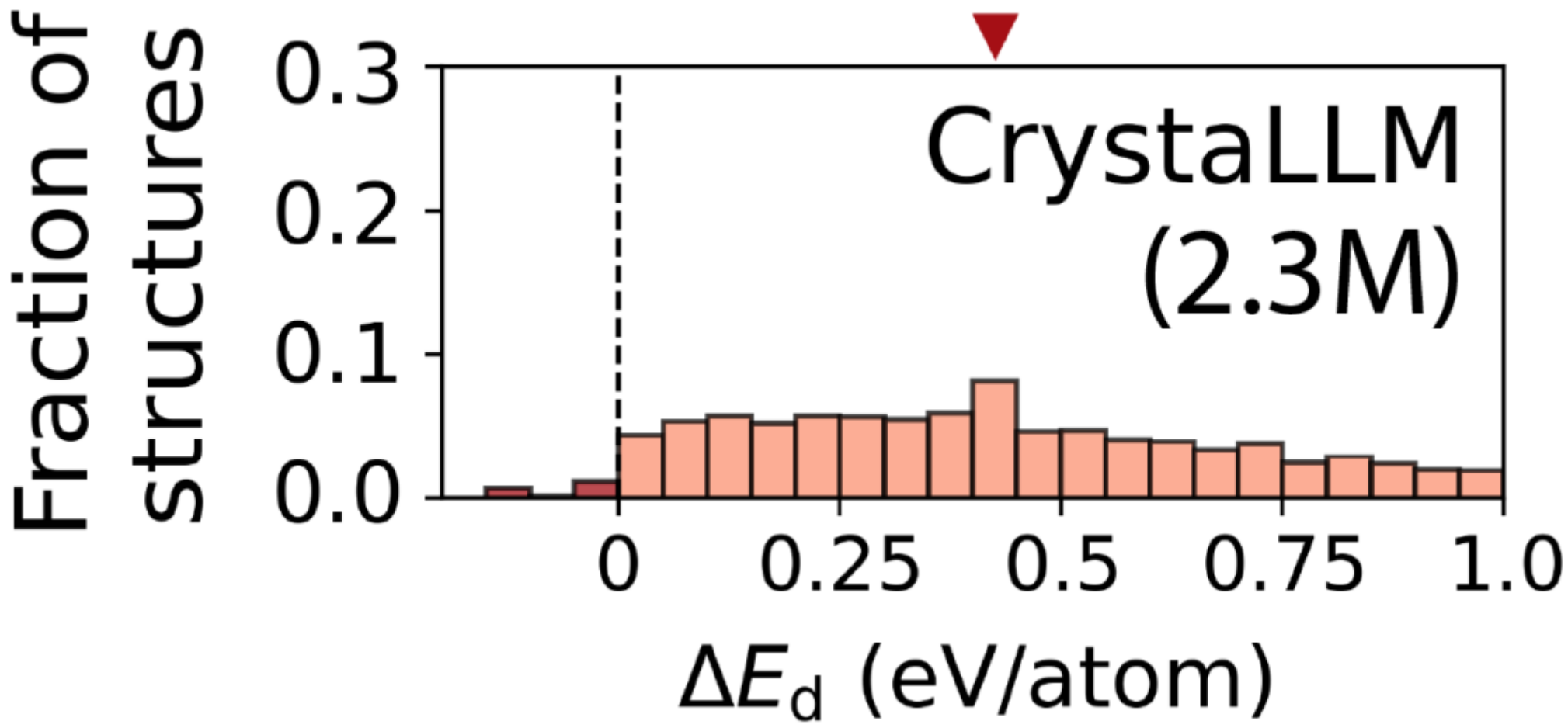

**Supplementary Figure 3:** Histograms showing density functional theory (DFT) computed decomposition energies of novel structures generated using a large version (200 million parameters) of CrystaLLM (from https://github.com/lantunes/CrystaLLM) trained on a dataset of 2.3 million structures, which is much larger than the MP-20 dataset (containing only 45,231 structures). The triangular marker indicates the median decomposition energy of generated structures.

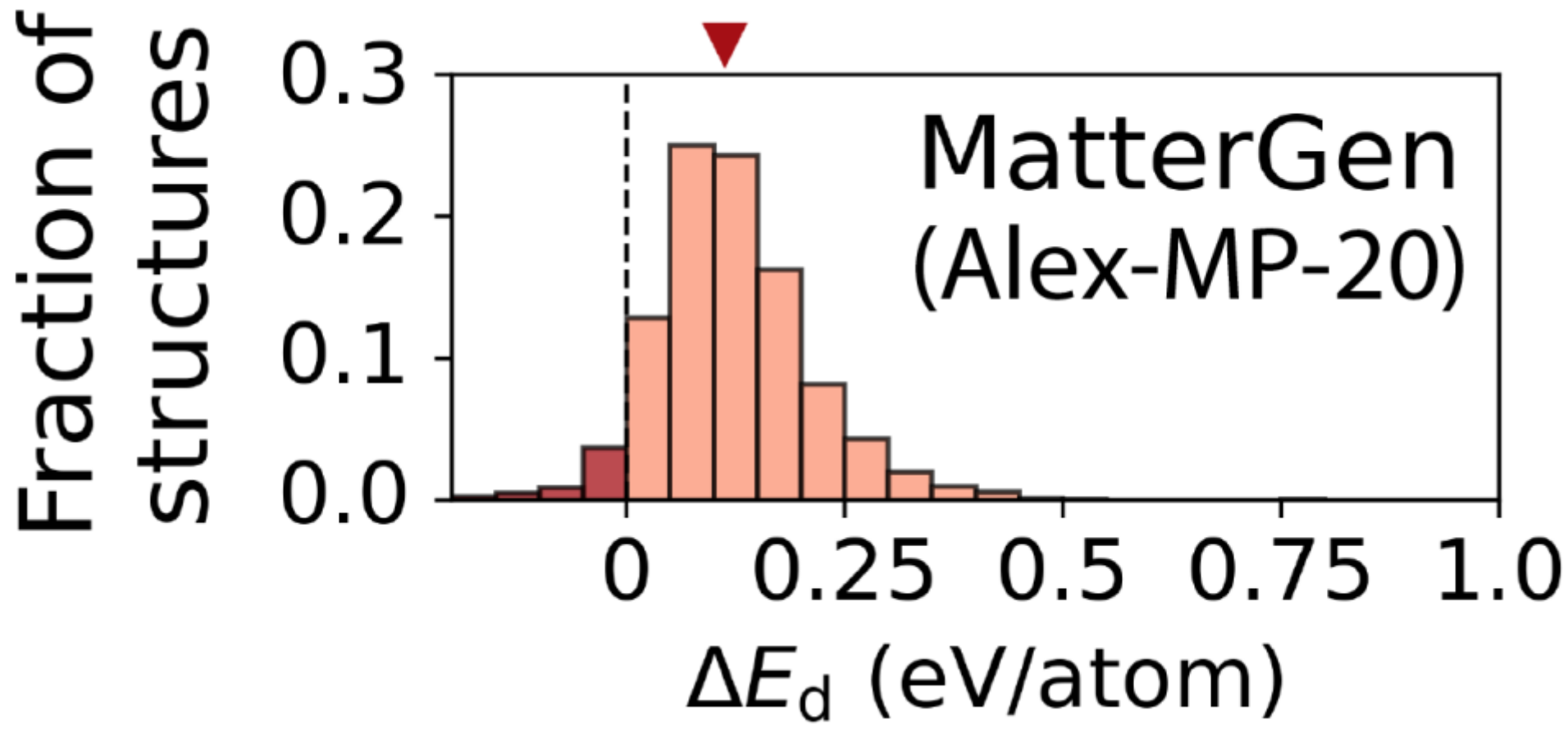

**Supplementary Figure 4:** Histograms showing density functional theory (DFT) computed decomposition energies of novel structures generated using a version of MatterGen that was trained on the Alex-MP-20 dataset (available at https://github.com/microsoft/mattergen). The triangular marker indicates the median decomposition energy of generated structures.

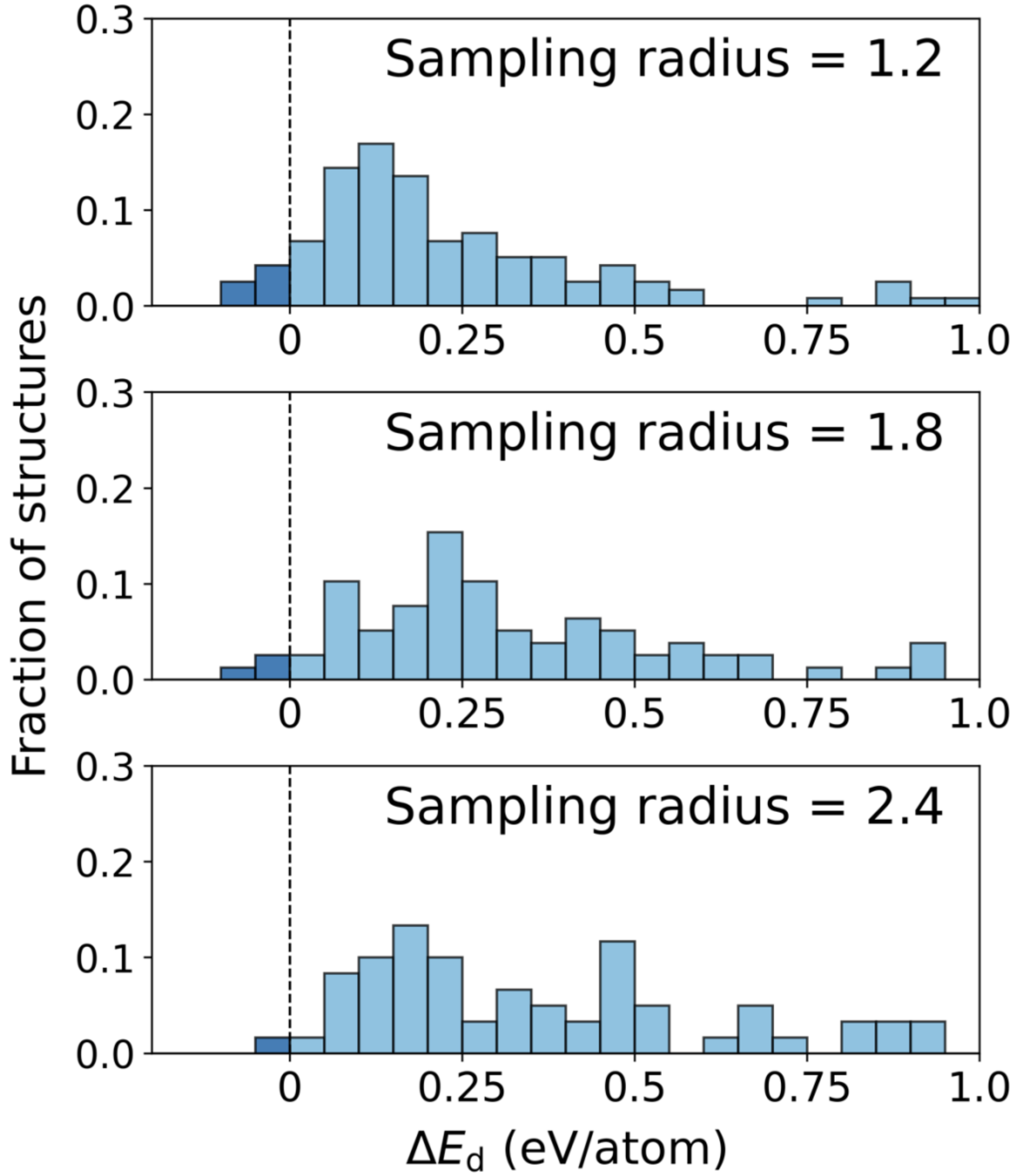

**Supplementary Figure 5:** Histograms showing density functional theory (DFT) computed decomposition energies of structures generated by FTCP as a function of sampling radius, which determines the distance from known materials in the latent space for sampling before decoding *via* the variational autoencoder. These statistics are based on 100 materials for each sampling radius.

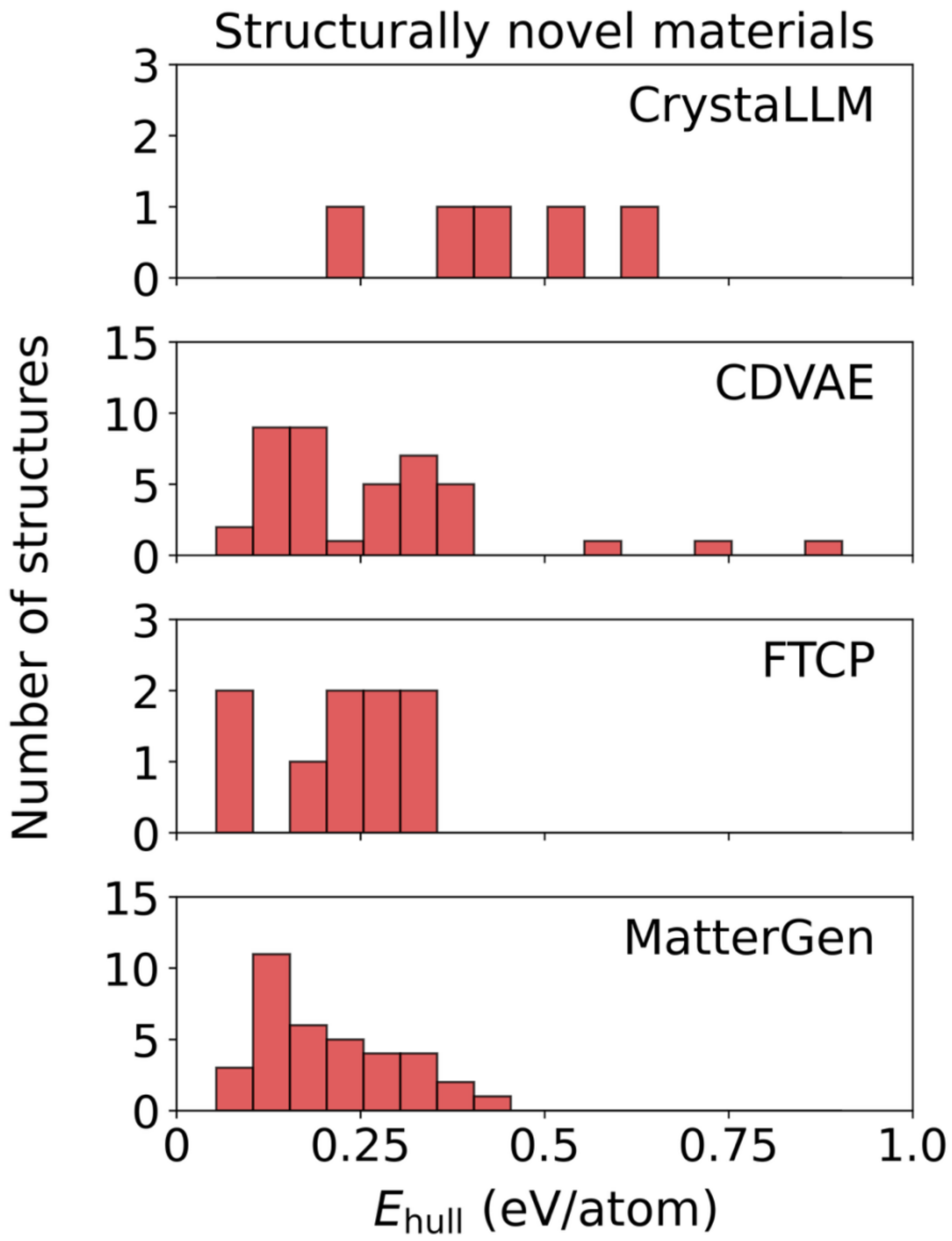

**Supplementary Figure 6:** Histograms showing density functional theory (DFT) computed energies above the convex hull for all the *structurally novel* compounds – meaning they cannot be mapped to known structure prototypes in the AFLOW database – created by four different generative models: CrystaLLM,[1] FTCP,[2] and CDVAE,[3] and MatterGen.[4]

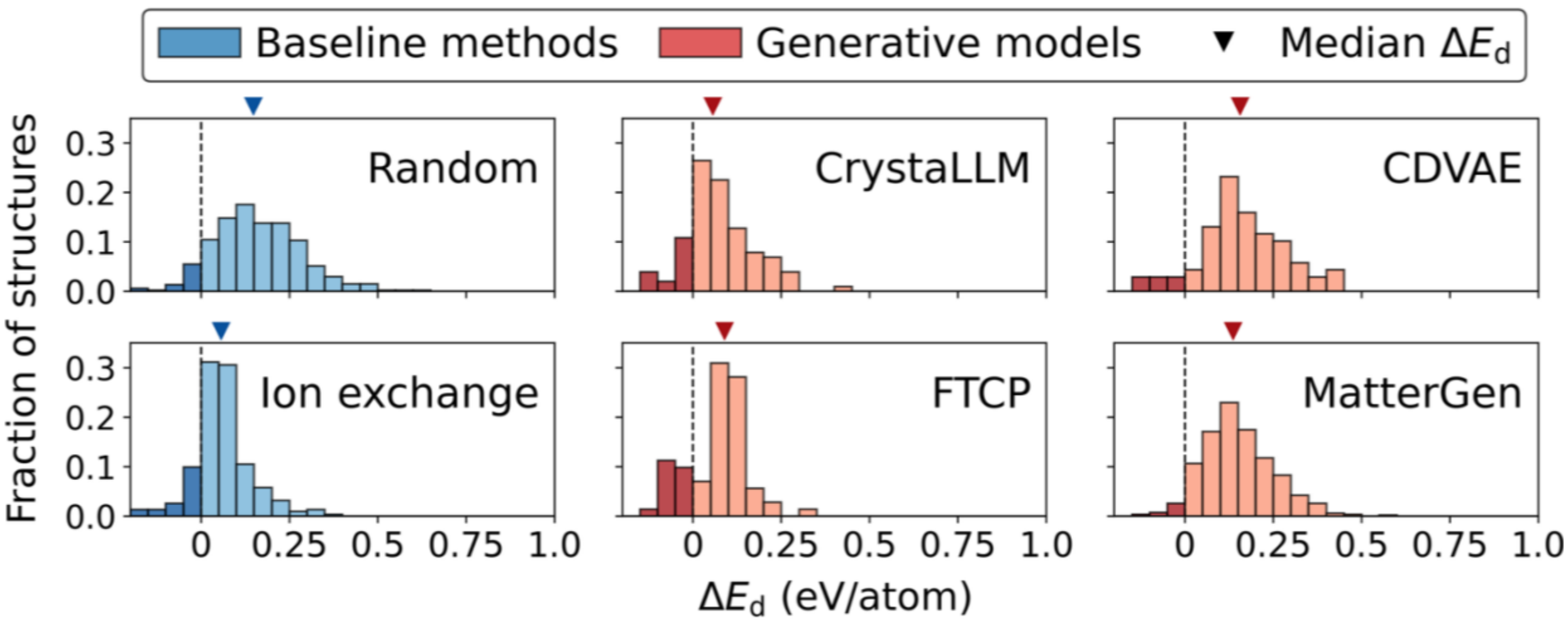

**Supplementary Figure 7:** Histograms showing DFT-computed decomposition energies ($\Delta E_\mathrm{d}$) for structures that were filtered by CHGNet-predicted stability[5]. These include structures generated by two baseline methods and four generative models. The left column (colored blue) contains results from the baseline methods: random enumeration and ion exchange. The right column (colored red) contains results from the generative model: CrystaLLM, FTCP, CDVAE, and MatterGen. These distributions are based on 300 structures from each method.

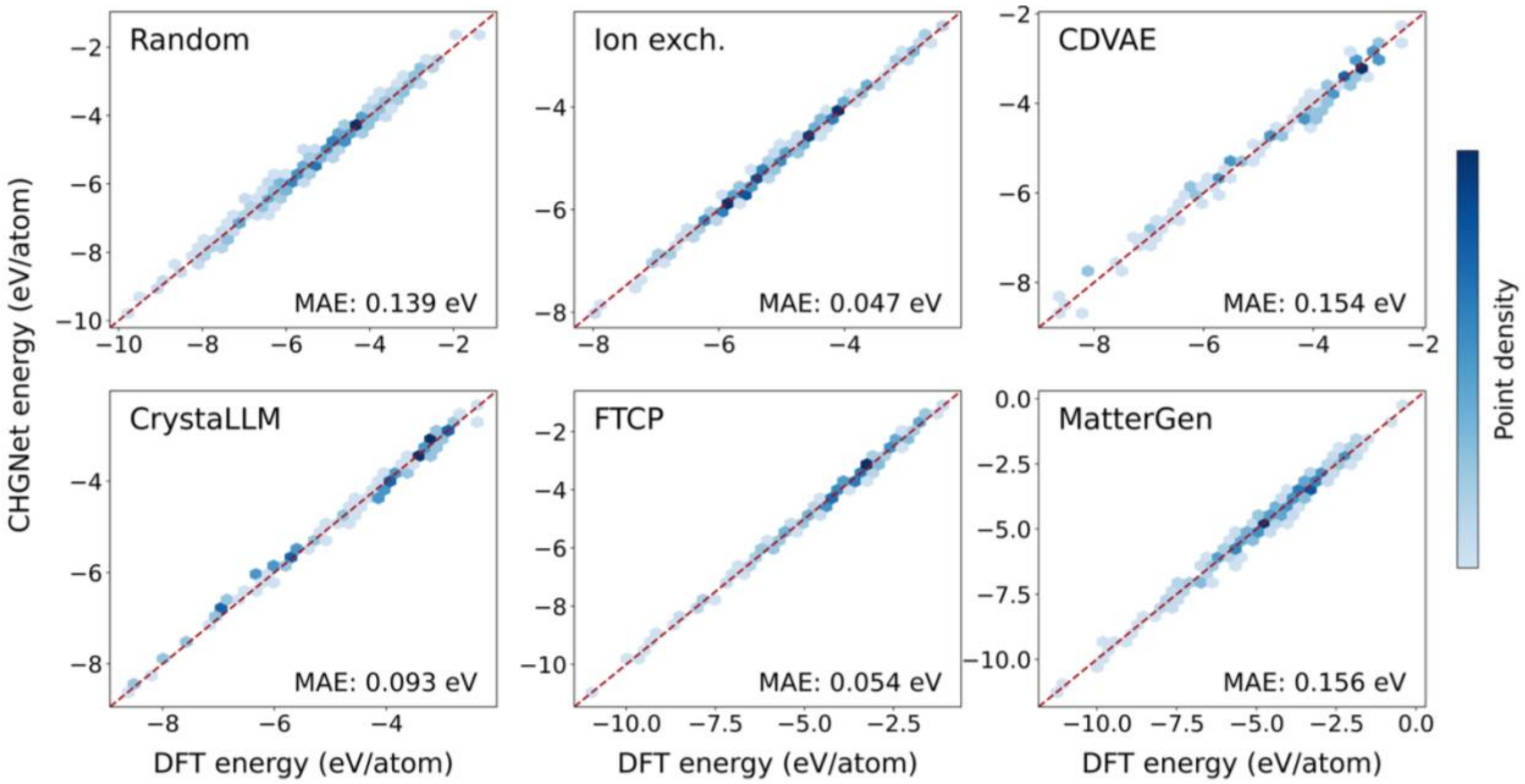

**Supplementary Figure 8:** Scatter plots comparing CHGNet-predicted with DFT-calculated energies of structures generated by two baseline methods (random enumeration and ion exchange) and four generative models: CrystaLLM, FTCP, CDVAE, and MatterGen. Both CHGNet and DFT were given the same starting structure and each relaxed it before computing a final energy. The plotted energies include corrections for anions and GGA/GGA+U mixing.[6] The large mean absolute errors (MAEs) are likely in part caused by the loose convergence criterion (forces were converged below 0.1 eV/Å) we used for CHGNet relaxations compared with DFT relaxations (0.03 eV/Å).

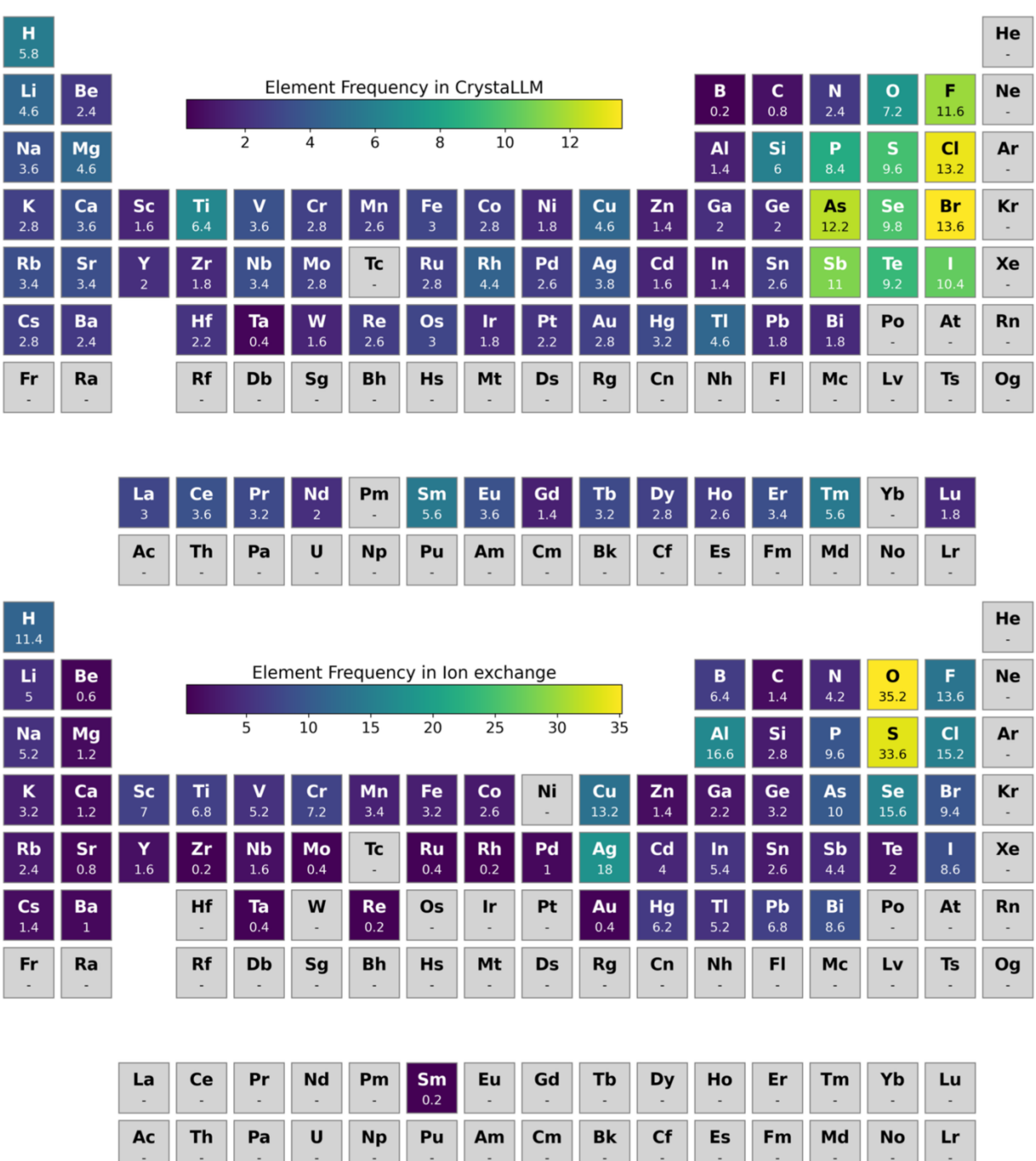
Element Frequency in CrystaLLM
2 4 6 8 10 12
H 5.8 He -
Li 4.6 Be 2.4 B 0.2 C 0.8 N 2.4 O 7.2 F 11.6 Ne -
Na 3.6 Mg 4.6 Al 1.4 Si 6 P 8.4 S 9.6 Cl 13.2 Ar -
K 2.8 Ca 3.6 Sc 1.6 Ti 6.4 V 3.6 Cr 2.8 Mn 2.6 Fe 3 Co 2.8 Ni 1.8 Cu 4.6 Zn 1.4 Ga 2 Ge 2 As 12.2 Se 9.8 Br 13.6 Kr -
Rb 3.4 Sr 3.4 Y 2 Zr 1.8 Nb 3.4 Mo 2.8 Tc - Ru 2.8 Rh 4.4 Pd 2.6 Ag 3.8 Cd 1.6 In 1.4 Sn 2.6 Sb 11 Te 9.2 I 10.4 Xe -
Cs 2.8 Ba 2.4 Hf 2.2 Ta 0.4 W 1.6 Re 2.6 Os 3 Ir 1.8 Pt 2.2 Au 2.8 Hg 3.2 Tl 4.6 Pb 1.8 Bi 1.8 Po - At - Rn -
Fr - Ra - Rf - Db - Sg - Bh - Hs - Mt - Ds - Rg - Cn - Nh - Fl - Mc - Lv - Ts - Og -
La 3 Ce 3.6 Pr 3.2 Nd 2 Pm - Sm 5.6 Eu 3.6 Gd 1.4 Tb 3.2 Dy 2.8 Ho 2.6 Er 3.4 Tm 5.6 Yb - Lu 1.8
Ac - Th - Pa - U - Np - Pu - Am - Cm - Bk - Cf - Es - Fm - Md - No - Lr -
Element Frequency in Ion exchange
5 10 15 20 25 30 35
H 11.4 He -
Li 5 Be 0.6 B 6.4 C 1.4 N 4.2 O 35.2 F 13.6 Ne -
Na 5.2 Mg 1.2 Al 16.6 Si 2.8 P 9.6 S 33.6 Cl 15.2 Ar -
K 3.2 Ca 1.2 Sc 7 Ti 6.8 V 5.2 Cr 7.2 Mn 3.4 Fe 3.2 Co 2.6 Ni - Cu 13.2 Zn 1.4 Ga 2.2 Ge 3.2 As 10 Se 15.6 Br 9.4 Kr -
Rb 2.4 Sr 0.8 Y 1.6 Zr 0.2 Nb 1.6 Mo 0.4 Tc - Ru 0.4 Rh 0.2 Pd 1 Ag 18 Cd 4 In 5.4 Sn 2.6 Sb 4.4 Te 2 I 8.6 Xe -
Cs 1.4 Ba 1 Hf - Ta 0.4 W - Re 0.2 Os - Ir - Pt - Au 0.4 Hg 6.2 Tl 5.2 Pb 6.8 Bi 8.6 Po - At - Rn -
Fr - Ra - Rf - Db - Sg - Bh - Hs - Mt - Ds - Rg - Cn - Nh - Fl - Mc - Lv - Ts - Og -
La - Ce - Pr - Nd - Pm - Sm 0.2 Eu - Gd - Tb - Dy - Ho - Er - Tm - Yb - Lu -
Ac - Th - Pa - U - Np - Pu - Am - Cm - Bk - Cf - Es - Fm - Md - No - Lr -

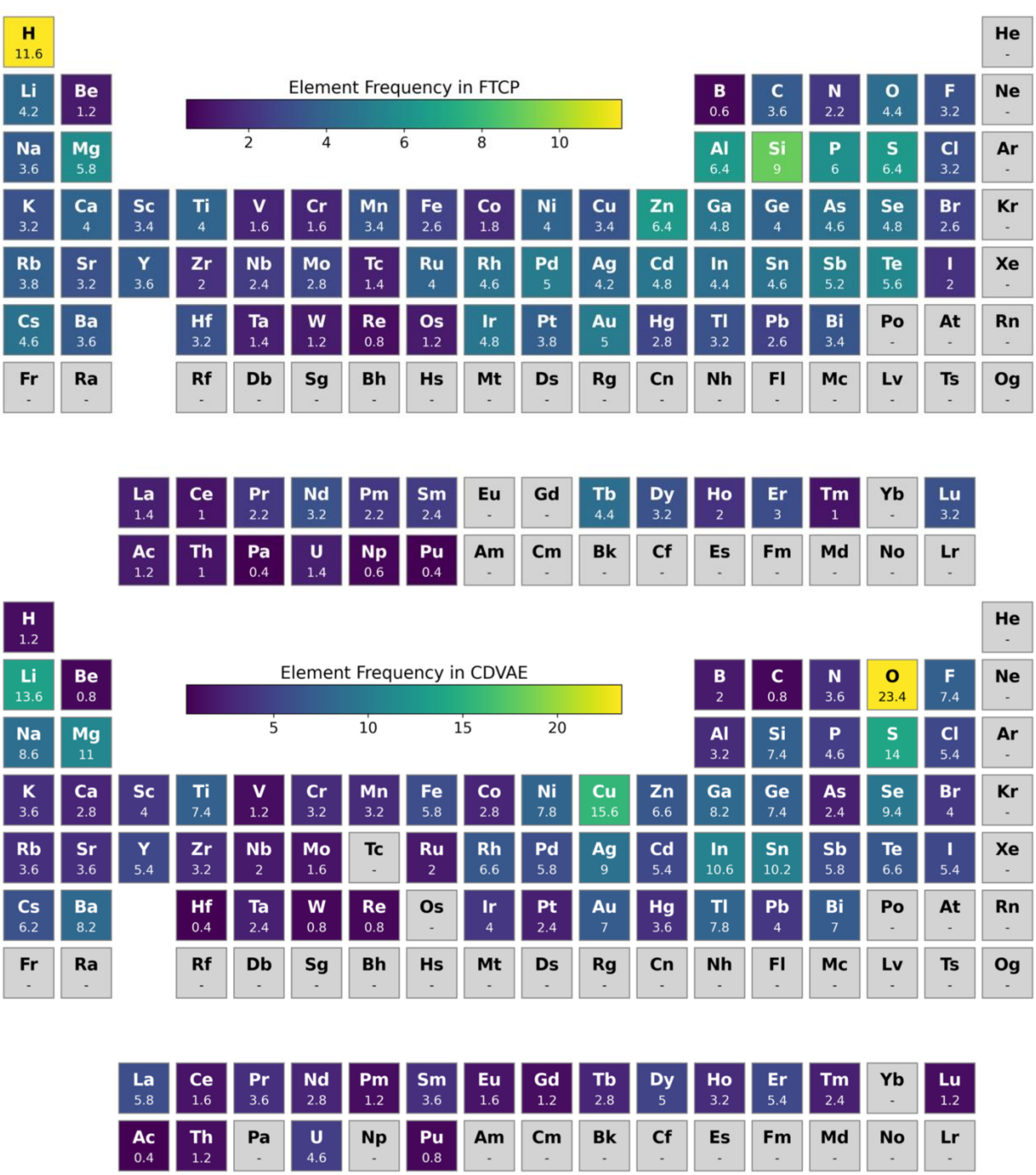
Element Frequency in FTCP
2 4 6 8 10
H 11.6 He -
Li 4.2 Be 1.2 B 0.6 C 3.6 N 2.2 O 4.4 F 3.2 Ne -
Na 3.6 Mg 5.8 Al 6.4 Si 9 P 6 S 6.4 Cl 3.2 Ar -
K 3.2 Ca 4 Sc 3.4 Ti 4 V 1.6 Cr 1.6 Mn 3.4 Fe 2.6 Co 1.8 Ni 4 Cu 3.4 Zn 6.4 Ga 4.8 Ge 4 As 4.6 Se 4.8 Br 2.6 Kr -
Rb 3.8 Sr 3.2 Y 3.6 Zr 2 Nb 2.4 Mo 2.8 Tc 1.4 Ru 4 Rh 4.6 Pd 5 Ag 4.2 Cd 4.8 In 4.4 Sn 4.6 Sb 5.2 Te 5.6 I 2 Xe -
Cs 4.6 Ba 3.6 Hf 3.2 Ta 1.4 W 1.2 Re 0.8 Os 1.2 Ir 4.8 Pt 3.8 Au 5 Hg 2.8 Tl 3.2 Pb 2.6 Bi 3.4 Po - At - Rn -
Fr - Ra - Rf - Db - Sg - Bh - Hs - Mt - Ds - Rg - Cn - Nh - Fl - Mc - Lv - Ts - Og -
La 1.4 Ce 1 Pr 2.2 Nd 3.2 Pm 2.2 Sm 2.4 Eu - Gd - Tb 4.4 Dy 3.2 Ho 2 Er 3 Tm 1 Yb - Lu 3.2
Ac 1.2 Th 1 Pa 0.4 U 1.4 Np 0.6 Pu 0.4 Am - Cm - Bk - Cf - Es - Fm - Md - No - Lr -
Element Frequency in CDVAE
5 10 15 20
H 1.2 He -
Li 13.6 Be 0.8 B 2 C 0.8 N 3.6 O 23.4 F 7.4 Ne -
Na 8.6 Mg 11 Al 3.2 Si 7.4 P 4.6 S 14 Cl 5.4 Ar -
K 3.6 Ca 2.8 Sc 4 Ti 7.4 V 1.2 Cr 3.2 Mn 3.2 Fe 5.8 Co 2.8 Ni 7.8 Cu 15.6 Zn 6.6 Ga 8.2 Ge 7.4 As 2.4 Se 9.4 Br 4 Kr -
Rb 3.6 Sr 3.6 Y 5.4 Zr 3.2 Nb 2 Mo 1.6 Tc - Ru 2 Rh 6.6 Pd 5.8 Ag 9 Cd 5.4 In 10.6 Sn 10.2 Sb 5.8 Te 6.6 I 5.4 Xe -
Cs 6.2 Ba 8.2 Hf 0.4 Ta 2.4 W 0.8 Re 0.8 Os - Ir 4 Pt 2.4 Au 7 Hg 3.6 Tl 7.8 Pb 4 Bi 7 Po - At - Rn -
Fr - Ra - Rf - Db - Sg - Bh - Hs - Mt - Ds - Rg - Cn - Nh - Fl - Mc - Lv - Ts - Og -
La 5.8 Ce 1.6 Pr 3.6 Nd 2.8 Pm 1.2 Sm 3.6 Eu 1.6 Gd 1.2 Tb 2.8 Dy 5 Ho 3.2 Er 5.4 Tm 2.4 Yb - Lu 1.2
Ac 0.4 Th 1.2 Pa - U 4.6 Np - Pu 0.8 Am - Cm - Bk - Cf - Es - Fm - Md - No - Lr -

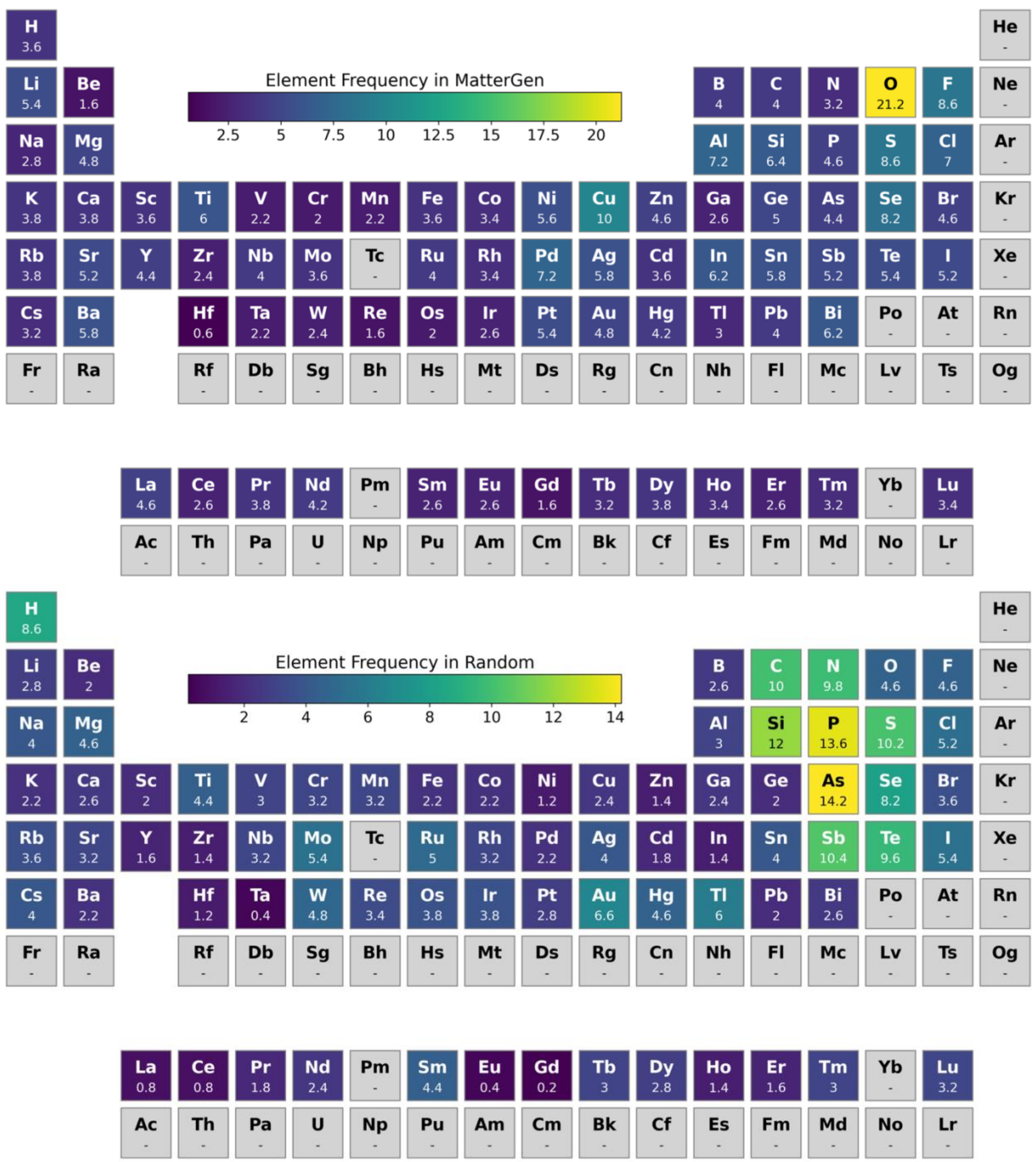

**Supplementary Figure 9:** Periodic table heatmaps representing the frequency of each element in the materials generated by random enumeration, ion exchange, CrystaLLM, CDVAE, FTCP, and MatterGen. The generative models were all trained on the MP-20 dataset.

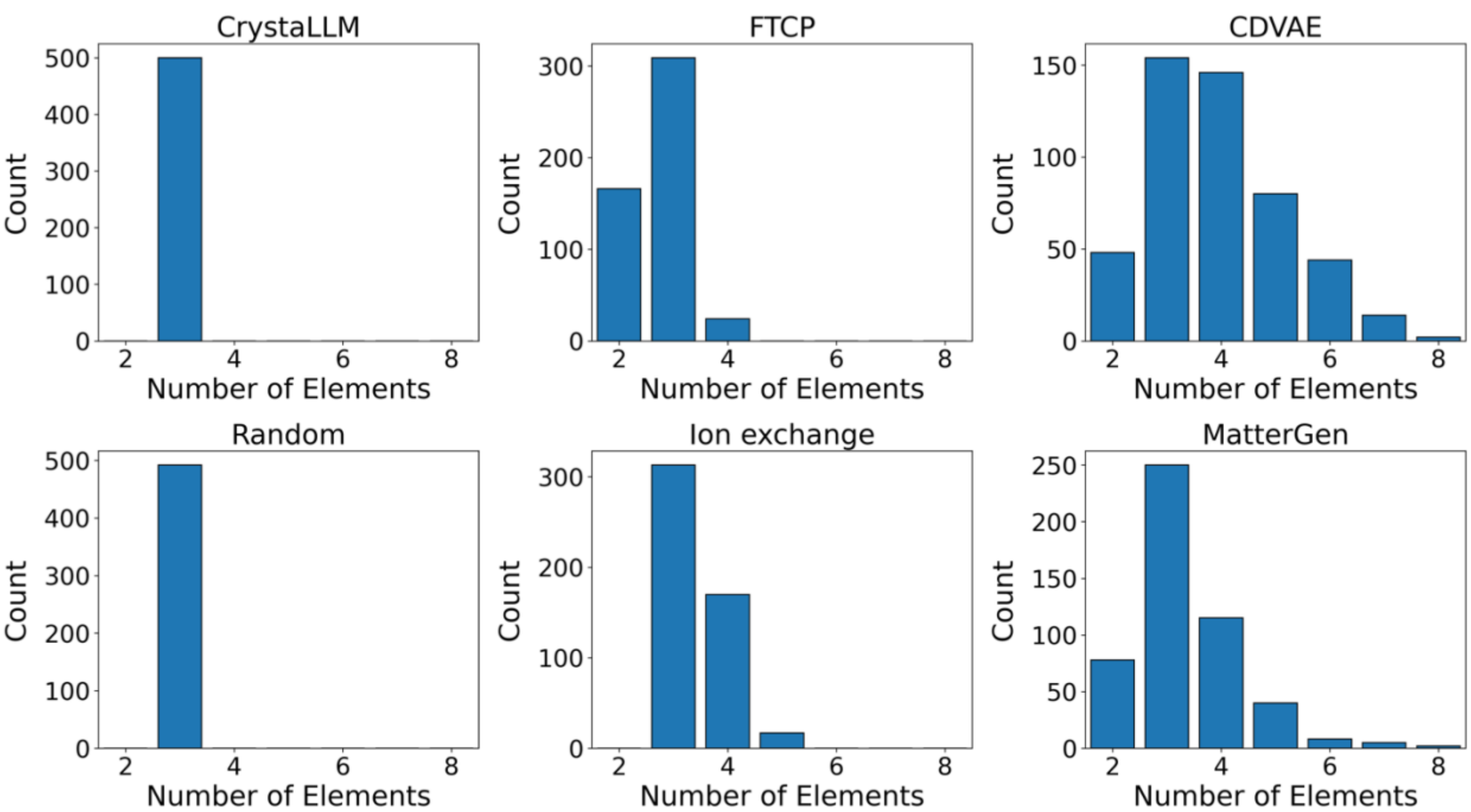

**Supplementary Figure 10:** Histograms showing distributions of the number of elements per novel material generated by each method.

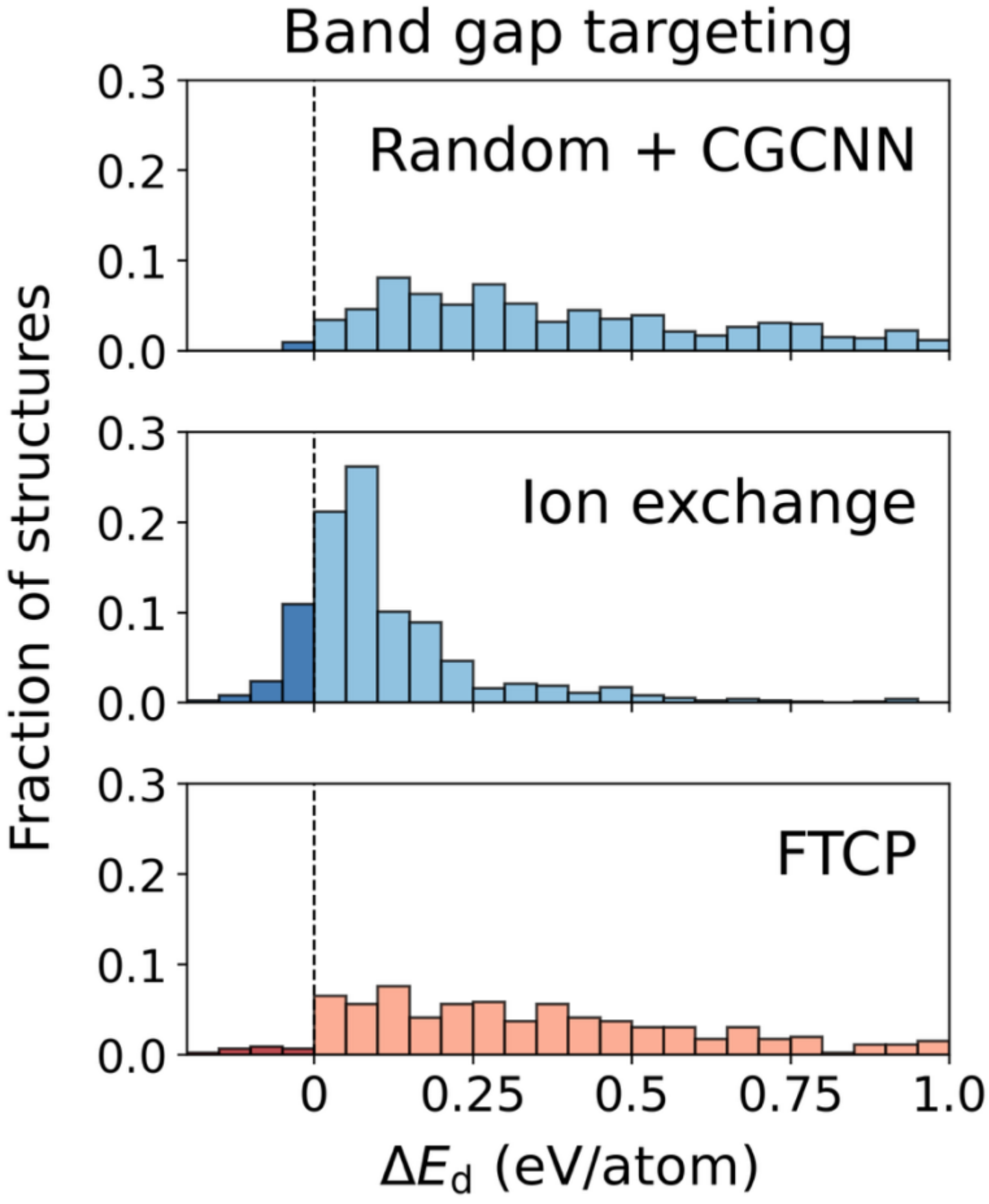

**Supplementary Figure 11:** Histograms showing DFT-computed decomposition energies of 500 novel materials generated by 1) using CGCNN to filter randomly enumerated materials, 2) data-driven ion exchange, and 3) a generative model, FTCP.[2] These approaches targeted materials with band gap near 3 eV, regardless of whether they are stable. Results from random enumeration are not included here since they follow the same distribution as shown in Figure 1 of the main text.

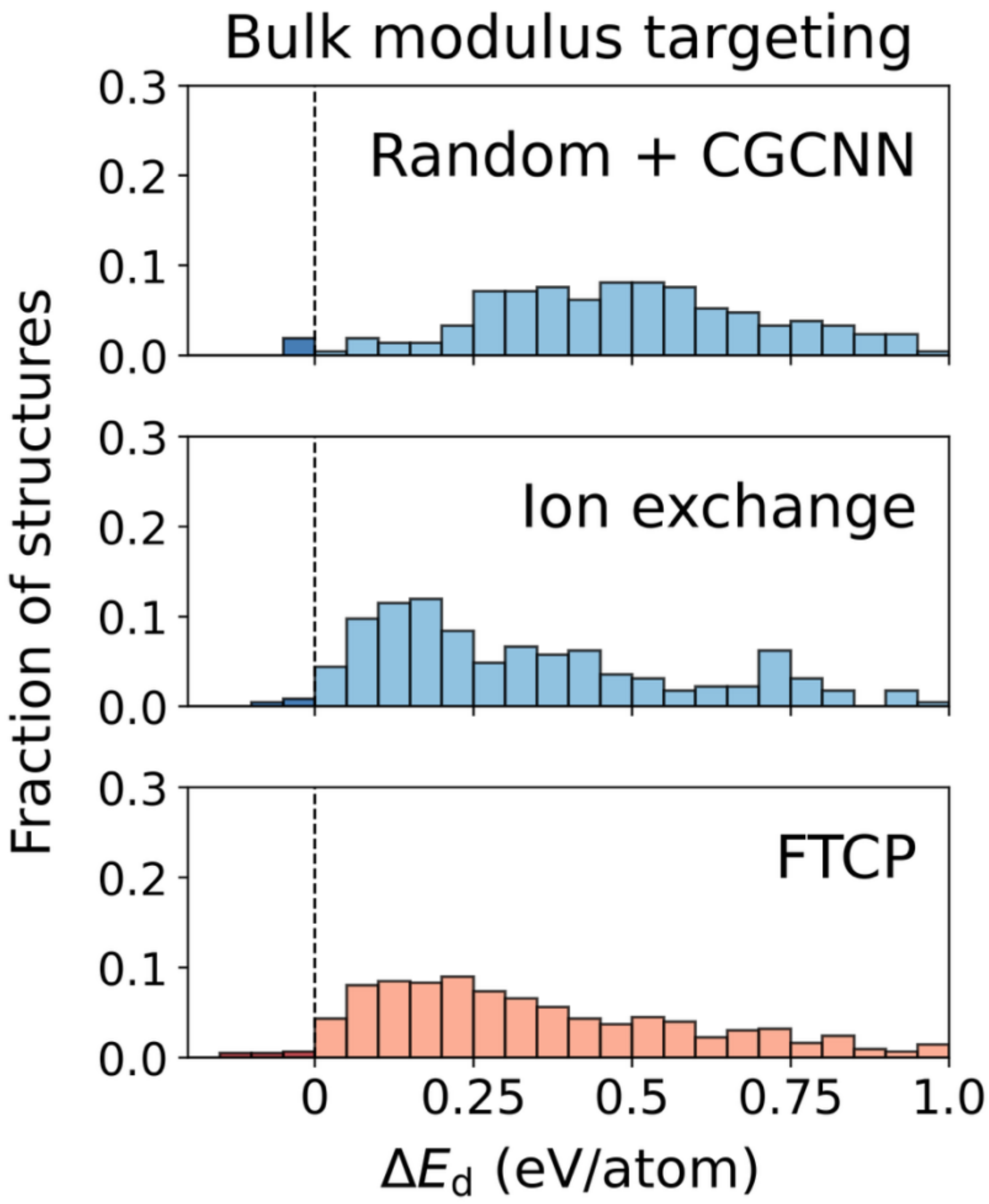

**Supplementary Figure 12:** Histograms showing DFT-computed decomposition energies of 500 novel materials generated by 1) using CGCNN[7] to filter randomly enumerated materials, 2) data-driven ion exchange, and 3) a generative model, FTCP.[2] These approaches targeted materials with a high bulk modulus, regardless of whether they are stable. Results from random enumeration are not included here since they follow the same distribution as shown in Figure 1 of the main text.

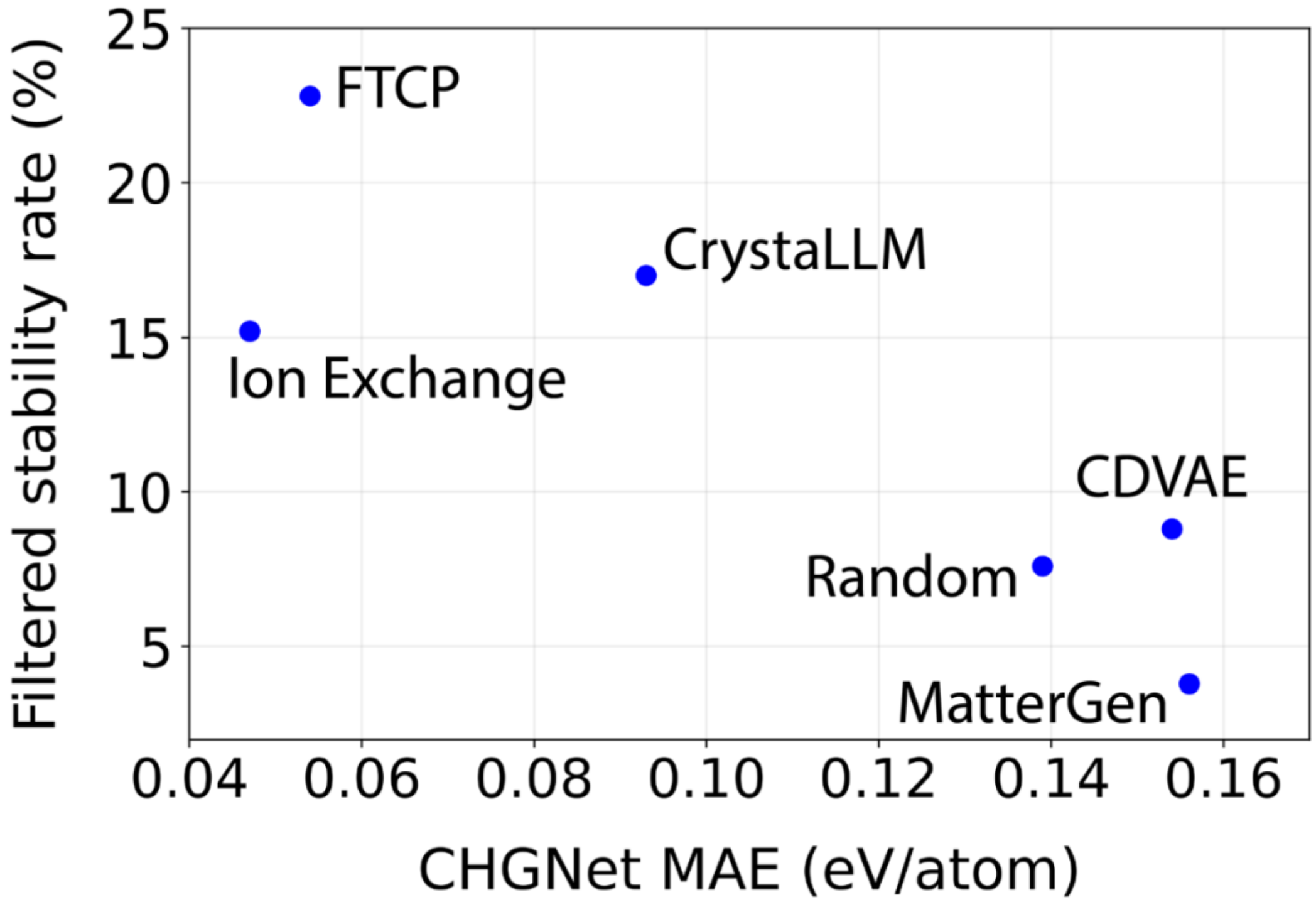

**Supplementary Figure 13:** Scatter plot showing correlation between filtered stability rate (y-axis) – the percentage of materials predicted as stable by CHGNet that remain stable in DFT – and mean absolute error (MAE) in CHGNet's predicted energy relative to DFT (x-axis).

## References

1. Antunes, L. M., Butler, K. T. & Grau-Crespo, R. Crystal Structure Generation with Autoregressive Large Language Modeling. Preprint at https://doi.org/10.48550/arXiv.2307.04340 (2024).

2. Ren, Z. *et al.* An invertible crystallographic representation for general inverse design of inorganic crystals with targeted properties. *Matter* **5**, 314–335 (2022).

3. Xie, T., Fu, X., Ganea, O.-E., Barzilay, R. & Jaakkola, T. Crystal Diffusion Variational Autoencoder for Periodic Material Generation. Preprint at https://doi.org/10.48550/arXiv.2110.06197 (2022).

4. Zeni, C. *et al.* A generative model for inorganic materials design. *Nature* **639**, 624–632 (2025).

5. B. Deng *et al.* CHGNet as a pretrained universal neural network potential for charge-informed atomistic modelling. *Nat. Mach. Intell.* **5**, 1031–1041 (2023).

6. Wang, A. *et al.* A framework for quantifying uncertainty in DFT energy corrections. *Sci Rep* **11**, 15496 (2021).

7. T. Xie & J. C. Grossman. Crystal Graph Convolutional Neural Networks for an Accurate and Interpretable Prediction of Material Properties. *Phys. Rev. Lett.* **120**, 145301 (2018).